\documentclass[a4paper,fleqn,usenatbib]{mnras}

\usepackage{newtxtext,newtxmath}

\usepackage[T1]{fontenc}
\usepackage{ae,aecompl}

\usepackage{graphicx}	
\usepackage{amsmath}	
\usepackage{amssymb}	

\usepackage{lscape}

\title{Most pseudo-bulges can be formed at later stages of major mergers}
\author[T. Sauvaget et al.]{T. Sauvaget$^{1}$\thanks{E-mail: tabatha.sauvaget@obspm.fr}, F. Hammer$^{1}$, M. Puech$^{1}$, Y. B. Yang$^{1}$, H. Flores$^{1}$, and M. Rodrigues$^{1}$
 \\
$^{1}$ GEPI, Observatoire de Paris, PSL Research University, CNRS, Place Jules Janssen, 92190 Meudon, France
}

\date{Accepted 2017 September 18. Received 2017 September 18; in original form 2017 June 29.}

\pubyear{2017}

\begin{document}
\label{firstpage}
\pagerange{\pageref{firstpage}--\pageref{lastpage}}
\maketitle

\begin{abstract}
Most giant spiral galaxies have pseudo or disk-like bulges that are considered to be the result of purely secular processes. This may challenge the hierarchical scenario predicting about one major merger per massive galaxy ($>$$3
\times 10^{10} M_{\sun}$) since the last $\sim$ 9 billion years. Here we verify whether or not the association between pseudo-bulges and secular processes is irrevocable. Using GADGET2 N-body/SPH simulations, we have conducted a systematic study of remnants of major mergers which progenitors have been selected (1) to follow the gas richness-look back time relationship, and (2) with a representative distribution of orbits and spins in a cosmological frame. 
Analyzing the surface-mass density profile of both nearby galaxies and merger remnants with two components, we find that most of them show pseudo-bulges or bar dominated centers. Even if some orbits lead to classical bulges just after the fusion, the contamination by the additional gas that gradually accumulates to the center and forming stars later on, leads to remnants apparently dominated by pseudo-bulges. We also found that simple SPH simulations should be sufficient to form realistic spiral galaxies as remnants of ancient gas-rich mergers without need for specifically tuned feedback conditions.
We then conclude that pseudo-bulges and bars in spiral galaxies are natural consequences of major mergers when they are realized in a cosmological context, i.e., with gas-rich progenitors as expected when selected in the distant Universe.
\end{abstract}




\section{Introduction}
\indent In the local universe, 80\% of galaxies more massive
than $10^{10} M_\odot$ show a significant bulge in their
central parts (i.e., assumed with a Bulge-to-Total flux ratios $B/T >$ 0.05,
\citealt{Fisher2011}; see also \citealt{Kormendy2010}).
Bulges can be separated into two broad types: classical bulges and
pseudo-bulges. Classical bulges exhibit properties resembling
elliptical galaxies, while pseudo-bulges are distinguished by their
more disk-like features including bars
\citep{Kormendy1993,Andredakis1994,Fisher2016}. A variety of
observational morphological and kinematic characteristics allow to
classify bulges into these two types
\citep{Kormendy2004,Athanassoula2005,Fisher2016}. Classical bulges are
 show a smooth distribution of stars,
are supported by velocity dispersion, and are strongly dominated by an old stellar
population. They also have a photometric radial profile similar to
mass surface density profile of elliptical galaxies with Sersic indices close to 4 (see
definition in Sect. 2). Pseudo-bulges may show younger stars, even active star formation,
and their dynamics is less supported by velocity
dispersion than classical bulges. Pseudo-bulges can result from the
evolution of bars, which translate into boxy/peanut isophotal shapes
when viewed edge-on. Gas accumulated at the center of a galaxy as a
result of the bar gravitational torques can also result in inner disk
structures called disc-like pseudo-bulges. These share many properties
of disk photometric profiles close to exponential with possible substructures such as spirals,
star-forming knots or dust lanes \citep{Athanassoula2005}. All pseudo-bulges share the properties that they can be identified as extra light
in the central part above the exponential profile of the outer disk. In this paper, we will simply refer as ''pseudo-bulges'' all such structures in central parts of spiral galaxies that
are non-classical bulges.

Classical bulges are considered to be the result of galaxy mergers,
while pseudo-bulges and bulgeless galaxies are believed to be the product of secular evolution
\citep{Kormendy2004,Athanassoula2005}. Indeed, when there is a limited
amount of gas in the progenitors of major mergers, the violent
relaxation of stars in the central parts of the remnant results in a
density profile with a high Sersic index \citep{Lynden-Bell1967}.
Conversely, secular processes correspond to smoother dynamical
processes in which gas is brought gradually to the center, resulting
in pseudo-bulges \citep{Athanassoula2005, Debattista2006, Heller2007, Athanassoula2013}.

The fact that pseudo-bulges or bulgeless galaxies are present in two thirds of
large nearby spiral galaxies \citep{Weinzirl2009,Kormendy2010},
seems to be in tension with the $\Lambda$CDM cosmological model
predicting several merger episodes per such galaxy
\citep{Stewart2008,Hopkins2010}. Indeed, is has been shown that more
than 50\% of present-day galaxies with stellar masses larger than $3
\times 10^{10} M_{\sun}$ have experienced one major merger since $z
\sim 1.5$ \citep{Hammer2005,Hammer2009,Puech2012,Rodrigues2017} leading often to a disk-dominated remnant. In this context, it might seem
difficult to explain how a majority of local spiral galaxies could be
devoid of classical bulges.

Perhaps spiral galaxies could be merger remnants of sufficiently ancient events implying that their progenitors are high redshift galaxies, the latter being gas-rich galaxies \citep{Erb2008,Rodrigues2012}. 
During gas-rich mergers (gas fractions in the progenitors
$\geq 50\%$ at the fusion time), the large amount
of gas involved in the process can dampen down the violent relaxation process
otherwise taking place at the center of the remnant, and result in
much reduced classical bulge fraction \citep{Hammer2005, Robertson2006, Hopkins2009, Hammer2009}. 
At larger scales, the gas inherits
its angular momentum from the orbital momentum of the merger and can
then be redistributed into a thin disk
\citep{Barnes2002, Hopkins2010,Athanassoula2016}.
The subsequent virialization phase of the rebuilt disk can last
several Gyr \citep{Hopkins2009,Puech2012} during which instabilities
could develop and play a role in the formation of pseudo-bulges,
especially in presence of bars that can redistribute gas towards the
center. 

The goal of this paper is to investigate how pseudo-bulges
could form in the centers of disks rebuilt after gas-rich major mergers.
It also extends previous studies using N-body/SPH simulations of
gas-rich major mergers. \citet{Hopkins2010} investigated how the
remnant $B/T$ scales as a function of the progenitor masses, gas
fractions, mass ratio using a large library of binary gas-rich mergers
between spiral progenitors. They found that major mergers are a
dominant channel for bulge formation, and that (at first order) $B/T$
decreases when the gas fraction increases. \citet{Keselman2012}
presented seven simulations of gas-pure major mergers sampling a
restricted range of orbits and spin orientation between the
progenitors, and found that the inner structures formed are consistent
with expectations from pseudo-bulges.

\citet{Athanassoula2016} included a hot gaseous component in the
progenitor models and showed that the two types of bulges can coexist
in the remnants, with, on average, only $\sim$ 10 - 20\% of the
stellar mass of the remnants ending in the classical bulge. Their
study also highlight that contrary to what is generally assumed in
simulations of isolated bar-forming galaxies, the bar is formed before
the thin disk is completely formed.
In this paper, we present a more systematic study of pseudo-bulge
formation using a library of 12 gas-rich major mergers hydrodynamical
simulations sampling all possible relative orientations (see \S3). In
\S4, we discuss the results in term of pseudo-bulge formation and
impact of bars, and compare observations and simulations. We begin
this study by revisiting the fraction of pseudo-bulges in local
galaxies using two complementary samples in \S2.

Throughout this paper, magnitudes are quoted in the $AB$ system and
stellar masses are estimated using a diet Salpeter IMF.

\section{pseudo-bulges in local spiral galaxies}

In this section we revisit the fraction of pseudo-bulge galaxies in
the local Universe using two complementary samples. The
\citet{Kormendy2010} is based on a very local Volume, within 8 Mpc. One one hand, it
therefore offers the finest spatial resolution, particularly
well-suited to distinguish pseudo from classical bulges. On the other hand, it
is limited by small statistics (only 16 spiral galaxies in total, see \S2.2), and it is likely not representative of the galaxy mass function since the \citet{Kormendy2010} restrict themselves to a
nearest volume that contains small groups of galaxies but not any denser environments. We therefore also investigated the fraction of
pseudo-bulges using the sample of 66 spiral and S0 galaxies from
\citet{Delgado2010}, who studied the morphology of a representative
sample of nearby galaxies selected in a much larger volume (at $0.02<z<0.03$, i.e., within 85 to 125 Mpc) from using bulge-to-disk
morphological decomposition (see \S2.3).

\subsection{Distinguishing pseudo from classical bulges}

The stellar distribution in the central parts can be characterized
using the Sersic index $n$ \citep{Sersic1963, Sersic1968}, which is
defined as follows :

\begin{equation}
I(r)=I_eexp \left ( -b_n\left[ \left( \frac{R}{R_e} \right )^{1/n} - 1\right] \right )
\end{equation}

\bigskip

\noindent in which $I(r)$ is the stellar luminosity profile, $R_e$ is the half-light radius of the bulge ($R_e = (b_n)^nr_0$ with
$b_n \approx 2.17n - 0.355$, $r_0$ the scale length of the bulge, \citealt{Fisher2008, Fisher2010}), and $I_e$ is the luminosity at $R_e$.  It is generally accepted that pseudo-bulges can be identified as
 those having $n <$ 2, while classical bulges show $n \ge$ 2
 \citep{Fisher2008}. Pure disks can be described by exponential disks with $n = 1$, 
 in the following we call $h$ the scale length of the stellar disk and 
 $I_d$ is the luminosity at radius $h$.

Other criteria have been used to distinguish both types of bulges such
as, e.g., the velocity to velocity dispersion ratio of stars \citep{Athanassoula2005, Kormendy2010, Fisher2016}. However these require more expensive hence rarer
kinematic data, and \citet{Fabricius2012} demonstrated a good correlation between Sersic index and kinematics. Following this, we chose in this paper to identify pseudo-bulges
using the above criterion on $n$ to allow simpler comparison
between observations and simulations (see \S4).

\subsection{The Kormendy sample}

\citet{Kormendy2010} studied a volume limited sample in a sphere of 8
Mpc around the Sun. They considered all giant galaxies within this
volume with 150 $km.s^{-1}$ $ < V_{circ} < $ 250 $km.s^{-1}$, including the Milky
Way (MW) and M31. Using the local Tully-Fisher relation
\citep{Hammer2007, Pizagno2007}, this translates into a stellar mass
range $\sim 3.55\times10^{10} M_{\sun}$ $<$ $M_{stellar}$ $<$
1.3$\times10^{11}$$M_{\sun}$.

\begin{figure*}
{\centering \includegraphics[width=0.8\textwidth]{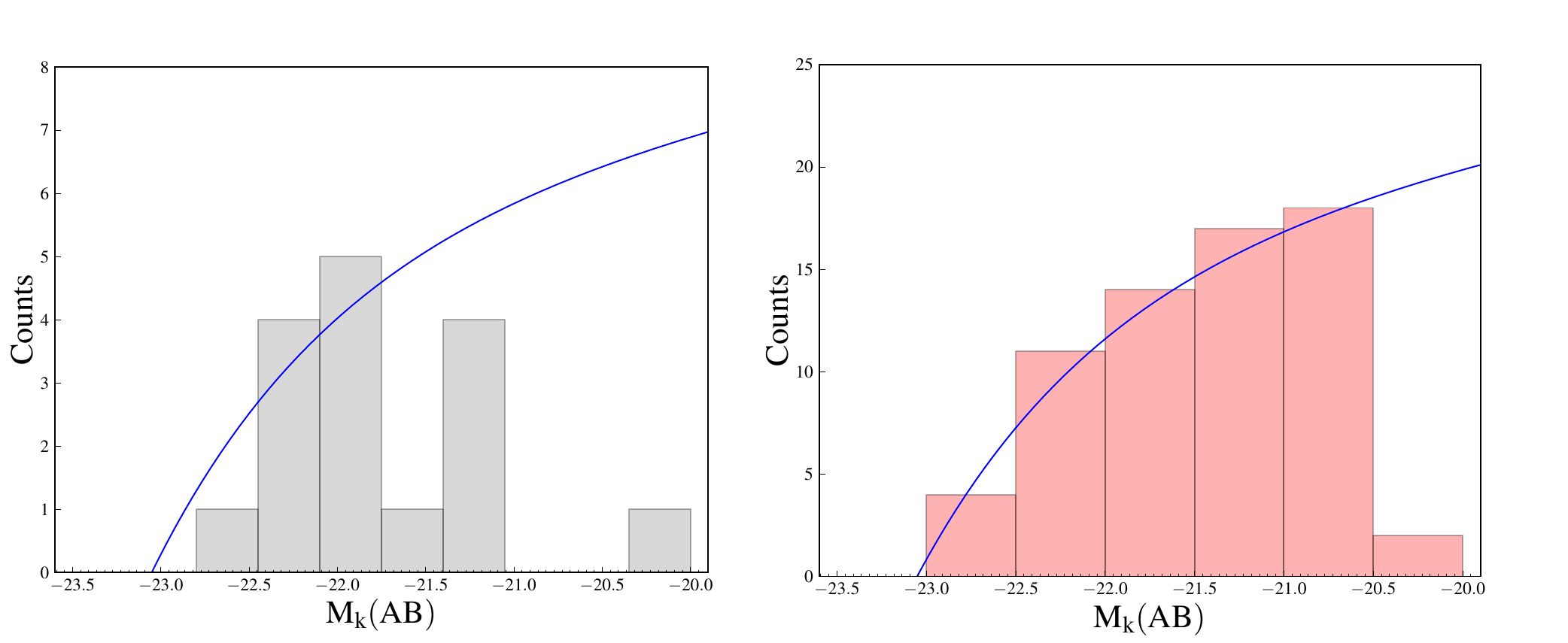}}
\caption{{\it Left panel :} Distribution of the absolute magnitudes for galaxies of the \citealt{Kormendy2010} sample, with $M_K$(AB)= -21.51 corresponding to $V_{circ}$ = 150 $km.s^{-1}$ from the Tully-Fisher relation \citep{Hammer2007}. {\it Right panel :} Magnitudes distribution of the 66 spiral galaxies selected from the \citealt{Delgado2010} sample. In both panels, the full (blue in the online version) line represents the K-band luminosity function from \citealt{Jones2006}. }
\label{Fig_mag_distribution}
	\end{figure*}

We checked for completeness using the local volume catalogue of
\citet{Karachentsev2013}, using the same \citet{Kormendy2010} criteria
on distance and velocity (using stellar mass as a substitute). One
spiral galaxy (Circinus) was removed from the sample because
\citet{Kormendy2010} used a rotation velocity of 155 $km.s^{-1}$, while this
value actually corresponds to $V_{max}$ rather than $V_{flat}$, which
is $\sim$ 100 $km.s^{-1}$ \citep{Curran2008}. We also updated the 2MASS
$K$-band magnitude of M31 and the MW used by \citet{Kormendy2010},
2MASS images being affected by sky subtraction residuals, 
to the more accurate values of
-24.51 (Vega) and -24.02 respectively, as described in
\citet{Hammer2007}. The resulting revisited sample of 16 spiral
galaxies\footnote{We note that in this sample, Maffei 2 is a barred
  spiral with a much lower surface brigthness which probably results
  in less accurate photometry compared to the other galaxies.}, including one SB0/a (NGC 2787), is
listed in Table \ref{Table_kormendy}. Figure \ref{Fig_mag_distribution}
shows the $K$-band magnitude histogram of the revisited sample.

\begin{figure}
\centering
\includegraphics[width=0.5\textwidth]{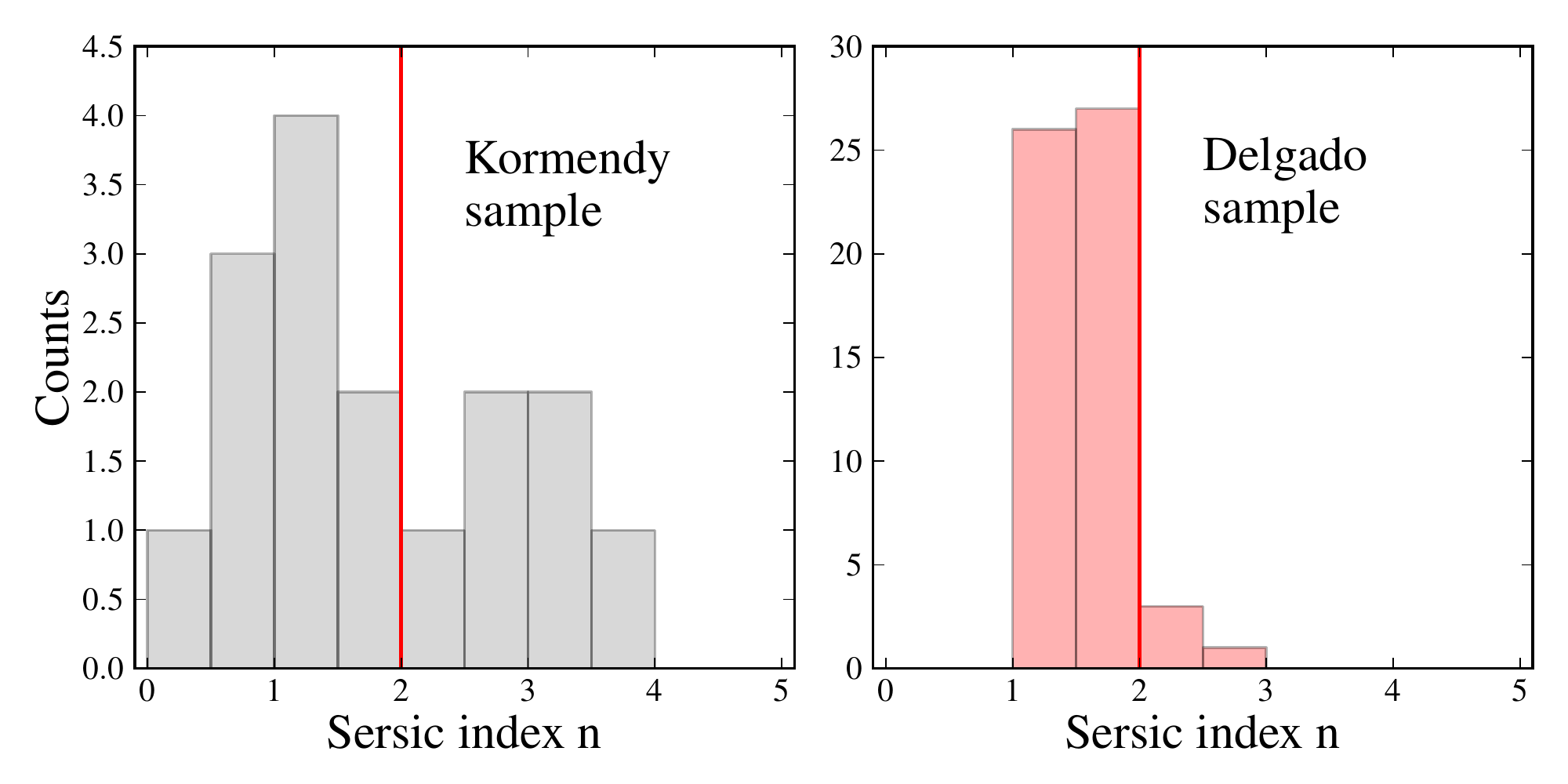} \\
  \caption{Distribution of the Sersic indices for the Kormendy sample ({\it left panel}) and for the Delgado sample ({\it right panel}).   
The red line (in the online version) corresponds to the cut at n = 2.}
\label{Fig_principale_histo}
\end{figure}

\citet{Kormendy2010} classified the bulges using a combination of
different criteria : Sersic index, $V/\sigma$, stellar formation from
molecular gas observations, and identification of disky structures in the center of galaxies such as
nuclear rings, bars, or nuclear disk. Amongst the 16 spiral galaxies
listed in Table \ref{Table_kormendy}, they found 62$\%$ of
pseudo-bulges (10/16), and 38$\%$ of classical bulges (6/16).
The distribution of Sersic indices is shown in Figure
\ref{Fig_principale_histo}. The fraction of galaxies with non-classical
bulges is  62\% $\pm$ 25\%  for a 1-$\sigma$ Poisson fluctuations in the parent sample.

\begin{table*}
  \caption{Spiral galaxies from the Kormendy sample. We have converted magnitudes from the Vega system (used in \citealt{Kormendy2010}) to the AB system by adding 1.85 for M31 and MW and 1.839 for the other galaxies. }

\begin{tabular}{lccccccc}
\hline
\hline
Name            & $M_K$ (AB)  &V ($km.s^{-1}$)    &$\sigma_v$       &     n        &      $\sigma_n$  &Ref for n value &Bulge type\\
\hline
\hline
NGC6946   & -21.77   &210   &10    &0.72  & 0.18 & \citet{Kormendy2010} &Pseudo\\
NGC5457    &-21.88   &210   &15    & 1.91  & -- & \citet{Kormendy2010} &Pseudo\\
IC342         & -21.39   &192   &3     & 1.80    &-- & \citet{Fisher2010} &Pseudo\\
NGC4945   & -21.37   &174   &10 & 1.30   &-- & \citet{Kormendy2010} &Pseudo\\
NGC5236   & -21.85   &180   &15 & 0.40   &-- & \citet{Fisher2010} &Pseudo\\
NGC5194  &  -22.10   &240   &20 & 0.50   &0.14 & \citet{Fisher2008, Fisher2010}  &Pseudo\\
NGC253   & -22.19   &210   &5    & 0.53   & -- & \citet{Simien1986} &Pseudo\\
Maffei2    &-21.16   &168   &20     & 3.00   &0.50 & \citet{Kormendy2010} &Pseudo\\
Galaxy      &-22.17     &220 &20     & --	& -- & \citet{Kormendy2010} &Pseudo\\
NGC4736    &-21.52   &181 &10   & 1.30  & 0.20 &\citet{Fisher2008, Fisher2010}  &Pseudo\\
NGC2683    & -21.28  &152 &5   &  2.50  &0.40 &\citet{Kormendy2010} &Classical\\
NGC2787   & -20.32   &220 &10  &  1.24  & 0.33 &\citet{Fisher2008} &Classical\\
NGC4826    &-21.87   &155 &5    &  3.00  &1.00 &\citet{Kormendy2010} &Classical \\
NGC4258   & -22.01   &208 &6    &  2.80  & 0.50 &\citet{Fisher2008, Fisher2010}  &Classical\\
M31            & -22.66    &250 &20  &   2.20  & 0.30 &\citet{Courteau2011} &Classical\\
M81           & -22.16     &240 &10  &3.80  & 0.10 &\citet{Fisher2008, Fisher2010}  &Classical\\
\hline
\hline
\end{tabular}
\label{Table_kormendy}
\end{table*}

\subsection{The Delgado-Serrano sample}
\label{section:delgado}
\citet{Delgado2010} studied the morphology of a mass-selected sample
of 116 galaxies with $M_{stellar} > 1.5\times10^{10}M_{\sun}$ at
$0.020 < z < 0.030$ from the Sloan Digital Sky Survey (henceforth
SDSS). This sample is representative of the K-band luminosity function
at these redshifts in this range of mass (see \citealt{Delgado2010}),
as shown in Figure \ref{Fig_mag_distribution}. From a two-dimensional
bulge+disk morphological decomposition in $R$ band using GALFIT
\citep{Peng2010}, they identified 101 spiral and S0 galaxies in this sample.
To avoid extremely edge-on galaxies and avoid any bias in analyzing
the inner regions, we considered in the following only the 66 spiral and S0
galaxies with a $b/a$ axis ratio larger than 0.35. \citet{Fisher2016}
argued that samples using SDSS images should not lie beyond 120
Mpc, which translates into z = 0.03 for a seeing of 1.4 arcsec. Due to its selection from the SDSS, the
\citet{Delgado2010} sample provides a good representation of the galaxy mass function, as well as a significantly larger sample than that of the Local Volume. 
Retrieving the bulge properties is however altered by the spatial 
resolution. \citet{Delgado2010} objective was to study the evolution of the Hubble
sequence as a function of redshift, but not to look in detail to the inner structures. 

We therefore
visually inspected the two-dimensional fits in these regions and found
occurrences of bars, rings, warps or tiny bulges not accounted for by the original fits. 
Because of the limited spatial resolution and signal-to-noise ratio (S/N) in these
structures, we repeated the disk+bulge morphological decomposition (see Figures in Appendix \ref{appen:degado}) using 1D average flux profiles constructed using the ellipse task from IRAF. 
For keeping an homogeneous treatment of both observations and simulations, we only fit two components per galaxy, assumed to be the bulge (with estimated Sersic n index) and the disk (with exponential profile).
It implies that the fitted bulge often contains a contribution from the bar, and the B/T ratio is more equivalent to a (bulge+bar)/total ratio. There are several motivations in doing so :
\begin{itemize}  
\item For both observations and simulations the spatial resolution definitively limits attempts of segregating the bar from the bulge;
\item In observed galaxies the bar signature may range from purely boxy bulges to large face-on viewed bars, depending on the inclination;
\item The flattening of the mass profile due to the softening is currently 0.2 kpc (2.8 times the softening radius, see Section~\ref{fit_simu}), which can be compared to FWHM/2=1.4 arcsec resolution, i.e., from 0.29 to 0.41 kpc for the observations.
\end{itemize}  
These similar spatial-resolution limitations may warrant a fair comparison between observations and simulations.  

Using the Sersic index criterion, bulges were then classified into
classical (n $\ge$ 2) vs. pseudo-bulges (n $<$ 2, see Table \ref{table_delgado}), and  found 88$\%$ $\pm$ 12$\%$
of pseudo-bulges or bulge-less galaxies (58/66), and
12$\%$ of classical bulges (8/66). 
 We find that classical bulge galaxies are mostly S0 galaxies in this analysis.
 Besides this, dust effects may potentially impact the morphological classification. To verify this, we further limit our sample to galaxies with $b/a$ $\ge$ 0.5 and 0.65, leading a fraction of 13$\%$ of classical bulges in the spiral population. It suggests that our result is not strongly biased by dust effects.
	
\subsection{Distribution of Sersic index in local spirals}
While the fractions of spiral galaxies with no classical bulges are
found to be consistent within 1-$\sigma$ uncertainties in both the
\citet{Delgado2010} and the \citet{Kormendy2010} samples, there might
be several reasons why this fraction might be significantly higher
in the former. While in the \citet{Delgado2010} sample bulges were
classified only on the basis of the Sersic index, \citet{Kormendy2010} used additional
criteria (see \S2.2). A more restrictive selection might naturally
explain a smaller fraction of non-classical bulges. In addition, the
\citet{Kormendy2010} study sampled more massive galaxies (compare the two panels in Figure 
\ref{Fig_mag_distribution}) for
which the fraction of classical bulges likely increases
\citep{Kormendy2010, Zahid2017}. While the Local
Volume sample is by construction complete (at least down to relatively
bright magnitudes) it is not necessarily representative of the local
luminosity function as the \citet{Delgado2010} sample does. However, the discrepancy between the two samples does not exceed the 1-$\sigma$ Poisson fluctuations
associated to the limited size of the two samples. Considering such
uncertainties, we can robustly conclude that significantly more than half of nearby
massive spirals are bulge-less or pseudo-bulged galaxies, i.e., do no
show classical bulges.

\section{Simulations of gas-rich major mergers}

We used a modified version of the public N-body/SPH code GADGET2
\citep{Springel2005}, in which star formation, SN feedback, and
cooling were modeled following \citet{Cox2006} as implemented by
\citet{Wang2012,Wang2015}. We adopted a constant feedback level 5$\times$
larger that median conditions as defined by \citet{Cox2006} in order
to preserve a large enough gas reservoir at fusion time. We verified
that changing the feedback value does not impact the profiles of the
remnants significantly. 
 A fiducial
softening radius of 73 pc (for the baryonic mass) was adopted along
with a resolution of 500\,000 particles. The core study of 12
gas-rich major mergers sampling the different possible geometric and
spin orientations between the two progenitors is described in \S3.1.
Mass profiles and disk+bulge morphological decompositions are detailed
in \S3.2. We checked that the resulting mass profiles are sufficiently
robust by repeating the same core study at high resolution (2
millions particles) as described in Appendix
\ref{appen:soft_and_res}, in which we also checked the impact of a
different softening radius (150 pc). All these parameters were found
to result in slight variations that do not impact significantly the
results presented in \S4.

\subsection{Gas-rich major merger models}
\label{section:orbits}
The two progenitors were modeled following \citet{Barnes2002}.
Briefly, the stellar mass of the main progenitor
(2.75$\times10^{10}M_{\sun}$) was scaled to approximately match the
mass ranges in the observed samples described in \S2. We adopted a
mass ratio between both progenitors of 3:1, which is the typical mass
ratio observed in distant galaxy gas-rich major mergers
\citep{Hammer2009}. Gas fractions in the progenitors were scaled to
typical values observed at $z\sim1.5$ \citep{Rodrigues2012}. This
corresponds to a look-back time of $\sim$ 9 Gyr, which matches the time
after which the remnant morphology was studied (see below). Both gas
and stars were distributed into pure thin exponential disks with a
gas extension 5 times more extended than the stellar disk. The stellar
disk sizes correspond to the typical disks at $z\sim 1.5$
\citep{Barden2005, Trujillo2006, Vanderwel2014}. A total of 20$\%$ in
baryonic mass (stars and cold gas) was assumed to approximately match
the average cosmological baryon density. Dark matter haloes were
modelled using a constant-density core profile as in
\citet{Barnes2002}. All components were initiated at rotation
velocities determined from the Tully-Fisher relation determined in
similar range of redshift and mass \citep{Puech2010}. All model
parameters are summarized in Table \ref{Table_param_simu}.

\begin{table*}
 \centering
  \caption{Parameters of the core study.}
  
\begin{tabular}{lr}
\hline
\hline
Parameter & Value\\
\hline
\hline
Stellar disk scale length of the first progenitor &3 kpc\\
Stellar disk thickness of the first progenitor & 0.2 kpc\\
Gas disk scale length of the first progenitor & 15 kpc\\
Stellar disk scale length of the second progenitor &1.73 kpc\\
Stellar disk thickness of the second progenitor & 0.115 kpc\\
Gas disk scale length of the second progenitor & 8.5 kpc\\
$a_{halo}$ of the first progenitor & 4 kpc\\
$a_{halo}$ of the second progenitor & 2.86 kpc\\
\hline
Total stellar mass of the first progenitor &     2.75 $\times10^{10}$$M_{\sun}$\\
Baryons fraction (bf) & 0.20       \\
Masses ratio (mr) & 3:1\\
Gas fraction of the first progenitor (cold gas) & 0.52\\
Gas fraction of the second progenitor (cold gas) & 0.72\\
Temperature of gas at the beginning & $10^4$ K\\
Rotation velocity of the first progenitor & 218 $km.s^{-1}$\\
Rotation velocity of the second progenitor & 145 $km.s^{-1}$\\
Feedback &constant (5 x median)\\
Pericenter (rp) & 16 kpc\\
Merger orbit & parabolic\\
\hline
Number of particles & 500 000 to 2 millions\\
Mass of star particles & 4$\times10^{5}$$M_{\sun}$ to $10^{5}$$M_{\sun}$ \\
Mass of gas particles &  1 x $M_{star}$\\
Mass of dark matter particles & 2 x $M_{star}$\\
Softening for star particles  &0.073 kpc\\
Softening for gas particles  & 0.073 kpc\\
Softening for dark matter particles  & 0.091 kpc\\
\hline
Stellar mass range of remnants within 15kpc &$3.99\times10^{10}$ - $5.47\times10^{10}$$M_{\sun}$\\
\hline
\hline
\end{tabular}
\label{Table_param_simu}
\end{table*}

\begin{figure*}
   \begin{minipage}[c]{.46\linewidth}

\begin{tabular}{lcccc}
\hline
\hline
Orbit & $i_1$ & $\omega_1$ & $i_2$ & $\omega_2$ \\
\hline
\hline
DIR-PROPRO &$0^{\circ}$ &$0^{\circ}$ &$71^{\circ}$ &$-30^{\circ}$\\
DIR-PRORET&$0^{\circ}$ &$0^{\circ}$ &$-109^{\circ}$ &$-30^{\circ}$\\
DIR-RETPRO &$180^{\circ}$ &$0^{\circ}$ &$71^{\circ}$ &$-30^{\circ}$\\
DIR-RETRET &$180^{\circ}$ &$0^{\circ}$ &$-109^{\circ}$ &$-30^{\circ}$\\
\hline
INC-PROPRO &$71^{\circ}$ &$30^{\circ}$ &$71^{\circ}$ &$30^{\circ}$\\
INC-PRORET &$71^{\circ}$ &$30^{\circ}$ &$-109^{\circ}$ &$30^{\circ}$\\
INC-RETPRO&$-109^{\circ}$ &$30^{\circ}$ &$71^{\circ}$ &$30^{\circ}$\\
INC-RETRET &$-109^{\circ}$ &$30^{\circ}$ &$-109^{\circ}$ &$30^{\circ}$\\
\hline
POLAR-PROPRO &$71^{\circ}$ &$-90^{\circ}$ &$71^{\circ}$ &$-90^{\circ}$\\
POLAR-PRORET &$71^{\circ}$ &$-90^{\circ}$ &$-109^{\circ}$ &$-90^{\circ}$\\
POLAR-RETPRO &$-109^{\circ}$ &$-90^{\circ}$ &$71^{\circ}$ &$-90^{\circ}$\\
POLAR-RETRET &$-109^{\circ}$ &$-90^{\circ}$ &$-109^{\circ}$ &$-90^{\circ}$\\

\hline
\hline

\end{tabular}

   \end{minipage} \hfill
   \begin{minipage}[c]{.51\linewidth}
\centering
\includegraphics[width=1.\textwidth]{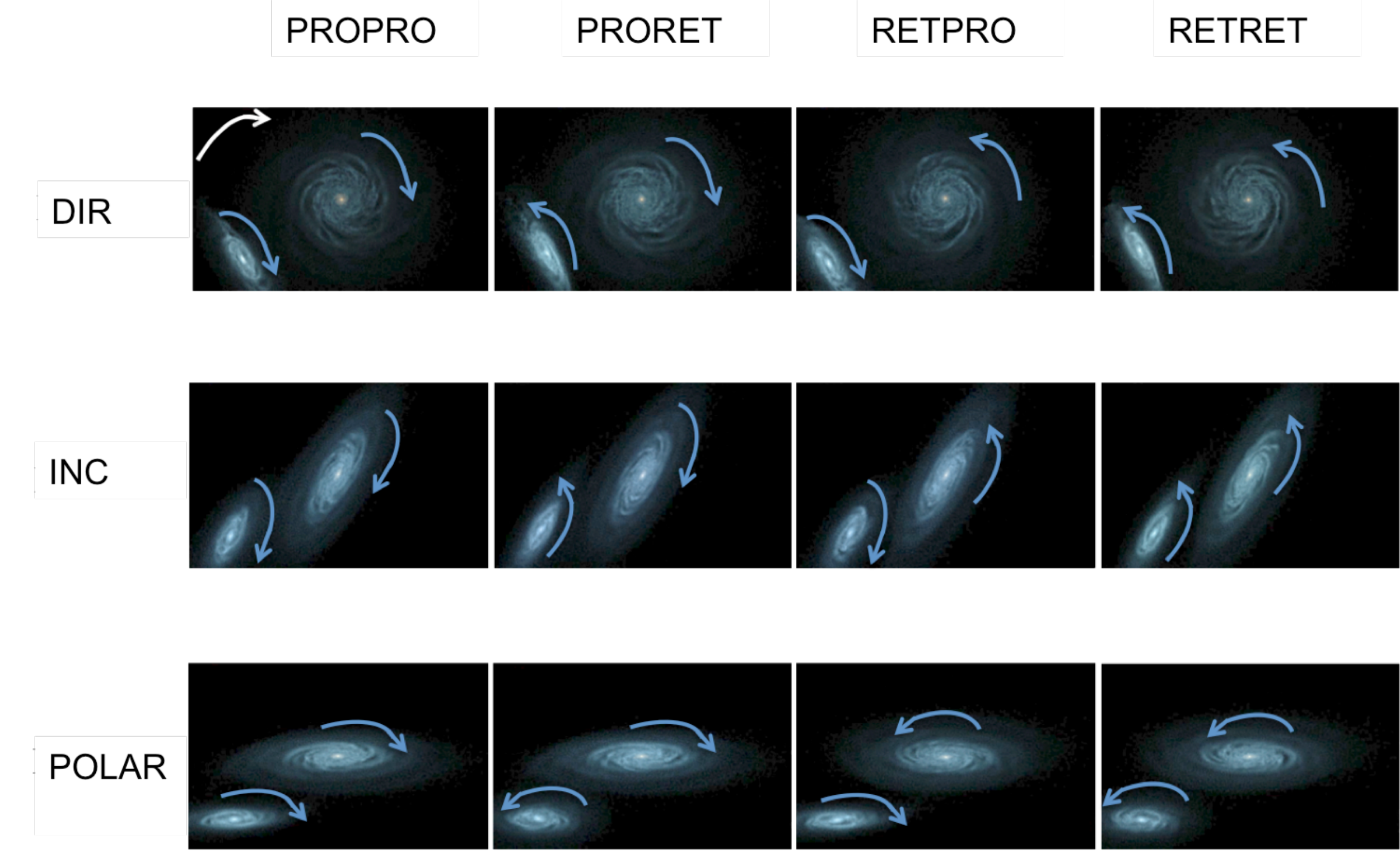}

   \end{minipage}
      
     \caption{{\it Left panel }: Orbital parameters of the two progenitors. i corresponds to the angle between the orbital plane and the spin plane of the galaxy, $\omega$ is the angle between the intersection axe (of the orbital plane and the spin plane) and the pericenter axe. {\it Right panel }: Snapshots of the first passage for the 12 simulations. The white arrow shows the direction of the merger orbit in the plan, blue arrows (in the online version) show the rotation of the progenitors (prograde or retrograde).}
     \label{Fig_orbit} 
\end{figure*}

The two progenitors were initiated on parabolic relative trajectories
following \citet{Barnes1992, Barnes1996, Barnes2002} (see Figure
\ref{Fig_orbit}). These sample the three general possible cases
describing the relative orientations between the main progenitor and
the orbital planes : direct orbits, in which the inclination between
the main progenitor disk and orbital planes $\sim 0^{\circ}$,
inclined, in which this angle $\sim 45^{\circ}$, and polar, with
$\sim 90^{\circ}$. To simplify the parameter space, the secondary progenitor was
chosen to have polar orbits but we checked that this as no significant
impact on the results. Conversely, spin orientations were found to
have significant impact so that we conducted a complete study of
relative spin orientations by sampling the 4 possible cases between
the two progenitors, i.e., prograde-prograde (PROPRO),
prograde-retrograde (PRORET), retrograde-prograde (RETPRO), and
retrograde-retrograde (RETRET). In this nomenclature, prograde means
that the progenitor rotation axis is aligned with the merger orbit,
while retrograde means an opposite rotation. This leads to a core study of
12 orbits (see Figure~\ref{Fig_orbit}). We note that these limited cases are nevertheless
representative in term of orientation of the main progenitor and
relative spins \citep{Barnes1992}. Indeed, this paper does not seek to
model specific observations but rather aims to construct a library
covering typical orbits that have significant
probability to happen in reality.

\subsection{Fitting bulges in merger remnants}
\label{fit_simu}
Simulations were run over about 12 Gyr to ensure that the remnants
are sufficiently virialized. Results were visually inspected to
measure characteristic times. The first passage and fusion times were
found to occur on average 1.5 and 3 Gyr after the beginning of the
simulation, respectively. It is broadly consistent with other
simulations producing disk-like remnants \citep{Cox2006,Cox2008,Brook2007,Lotz2008,Dutton2009}, with similar masses and gas fractions. 
We extracted snapshots after 9.3 Gyr of simulation, which
place the beginning of the simulations at $z \sim 1.5$, matching
the lookback-time at which gas fractions in the progenitors were
adopted from observations (see \S3.1), and providing self-consistent
simulations. This implies that remnants galaxies have virialized over
6.3 Gyr after fusion, therefore reaching sufficient equilibrium while
letting enough time for subsequent processes to develop during the latest stages of the merging.

From these snapshots, we centered (using the stellar barycenter within
30 kpc) and aligned the remnants to build face-on, stellar-mass surface
density maps. These profiles 
were then constructed (in $M_{\sun}/kpc^2$ units) within rings of
0.1 kpc and 1 kpc beyond a radius of 5 kpc to enhance signal-to-noise
ratio. They were compared to the profiles obtained
using the ellipse task within IRAF, which gives similar results. The 1D
profiles were used to decompose the remnants into an exponential disk
and an inner Sersic profile as described in \S2. Results are listed in
Appendix~\ref{Table_fit}.

\begin{figure}
\centering
\includegraphics[width=0.51\textwidth]{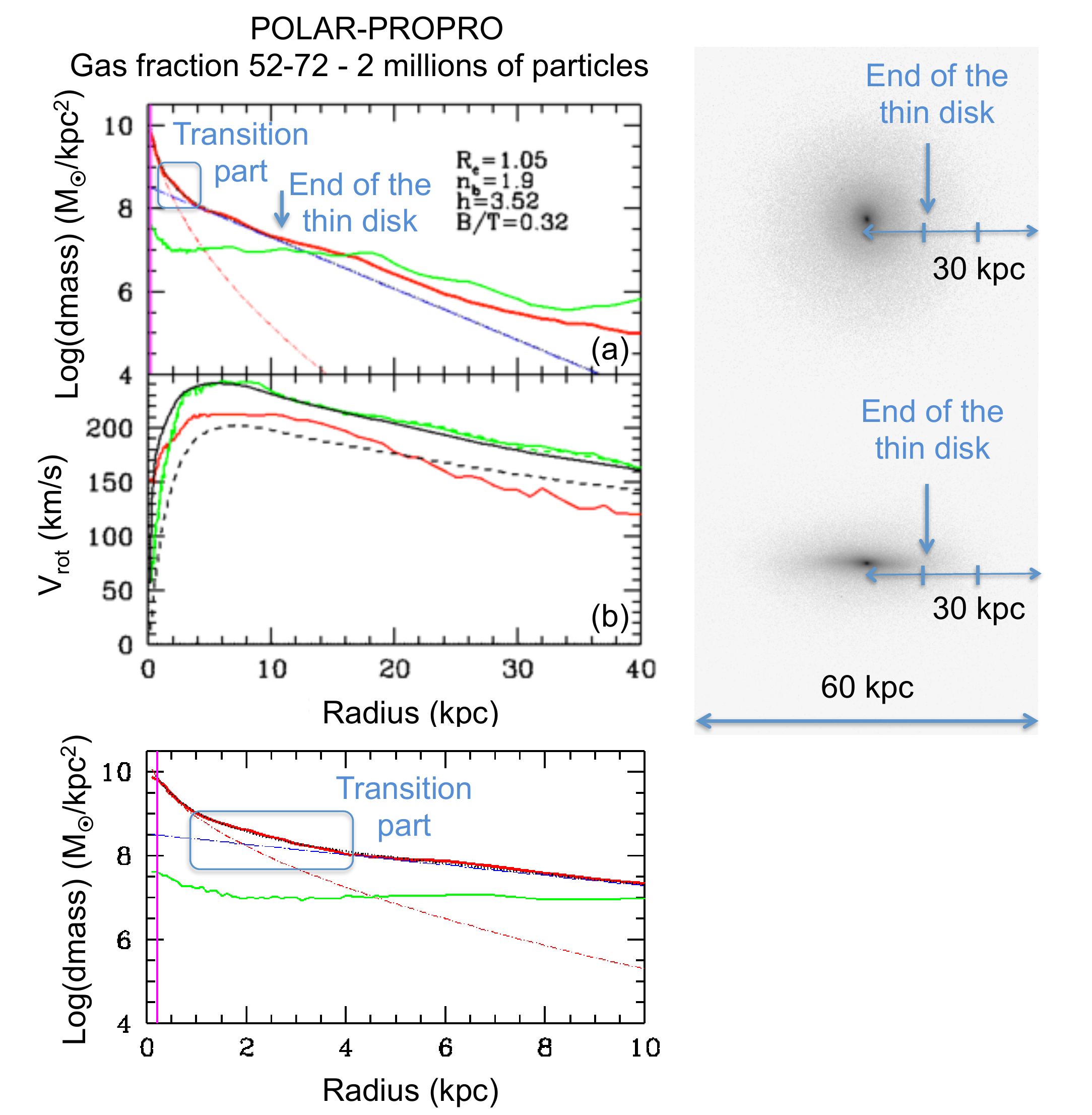} 
\caption{1D fit of mass profile for one simulation with the 2D analysis. {\it Top left panel (a) }: The thick red line (a color version of this figure is available in the online journal) corresponds to the stellar mass per $kpc^2$, the same profile for gas mass is shown in green. The dashed red line is the fit of the bulge component by a Sersic function, the exponential disk is fitted in the dashed blue line. The magenta line represents the softening radius (0.204 kpc). {\it Top left panel (b) }: Velocity curve of the analyzed galaxy, the red line corresponds to the stellar velocity curve, the green one to the gas velocity curve, which follows the velocity curve expected for the total mass (thick black line). {\it Bottom left panel }: Zoom of the 1D fit of mass profile. {\it Right panels }: 2D map of stellar mass on 60 kpc, face-on view on top panel, edge-on view on bottom panel. We can see a central bulge in black, the transition part in dark grey, the thin disk in grey and the outer disk in light grey.}
\label{Fig_example}
\end{figure}

Several difficulties could arise during such profile fitting. First, one
has to deal with a global degeneracy between the bulge and disk
parameters. Indeed, different sets of [$n$, $R_e$, $I_e$, $I_d$, $h$]
values can adjust the profile with similar resulting $\chi^2$, which
results in uncertainties on all fitted parameters. In particular, the
transition region between the bulge and the disk is often
degenerated between the two profiles (see Figure \ref{Fig_example}), in particular in presence of a
faint bar. The way to put the disk in relation to the
bulge could change the Sersic $n$ index of the bulge, resulting in 
uncertainties of $\pm 0.2$ in $n$, $\pm 0.1$ kpc in $R_e$, $\pm 0.3$ kpc in $h$, and $\pm 0.1$ in $B/T$. 
 Since we used the same method of 1D fitting than done for the Delgado sample analysis, 
we expect similar uncertainties for the latter than those estimated above.
In parallel to the fit, we used the 2D stellar mass density maps 
using both face-on and edge-on projections to correctly identify the
different structures as illustrated in Figure \ref{Fig_example}. The finite softening radius in the simulations result in a flattening
of the structures in the central regions below 2.8 times this radius in 
the center \citep{Springel2005}, just as the seeing does for observations. 
To avoid this problem, we have adjusted all the mass density profiles by excluding the
central parts affected by the softening radius. \\

\section{Simulation Results}

\subsection{Remnant galaxies are all spirals in the core study}
We observe in the simulations a behavior broadly consistent with
previous simulation studies (see references in Introduction). The two
progenitors orbits intersect at least two times (first and
second passage), while a bar often forms in the main progenitor at
the beginning of the simulation as a result of resonances between the
orbit of the secondary and the internal rotation \citep{Hopkins2009}. 
After a 3rd and in some cases a 4th passage, the fusion of the two nuclei takes place, on
average 3 Gyr after the beginning of the simulations. All along the
interaction gas and stellar particles are ejected into tidal tails in
which tidal dwarf galaxies may form. 
Many old stars end up in the central parts due to the violent relaxation, 
a small fraction of them in the thick disk and in the halo, 
while young stars converted from the gas form the thin disk and disky structures as bars or rings. 
In particular, we confirmed that gas-poor
progenitors may result in elliptical galaxies (see \S4.2), while gas-rich
major mergers reform thin disks.

\begin{figure*}
      \includegraphics[width=1.0\textwidth]{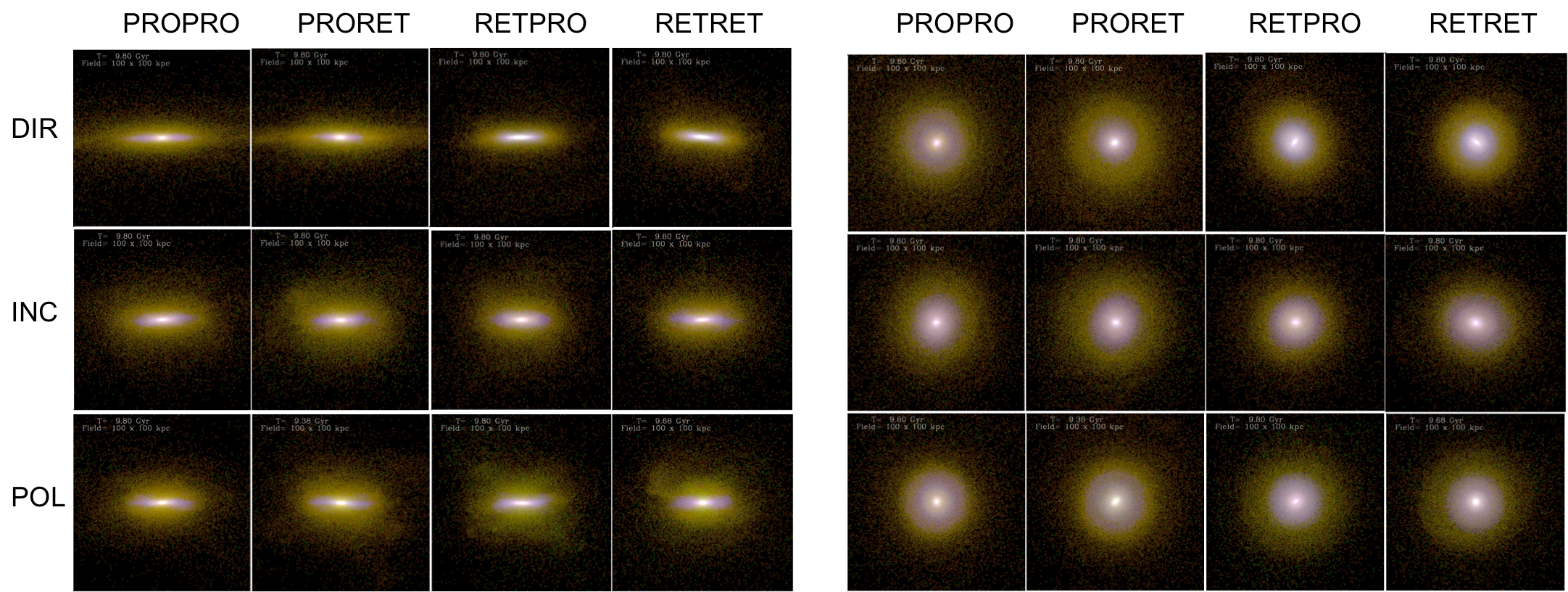}
     \caption{2D maps of stellar mass for the 12 remnant galaxies modelled with 2 millions particles. {\it Left} : Edge-on view, {\it Right} : Face-on view. DIR, INC and POLAR correspond to the three orbital planes of the merger. PROPRO, PRORET, RETPRO and RETRET correspond to the spin orientations of progenitors: prograde or retrograde compare to the direction of the orbit merger (see Section \ref{section:orbits} for more details). Old stars are seen in yellow (in the online version) and young stars are seen in blue (born since the fusion). One can see several components : bulges in the center, bars and thin disks in blue, double-disks, thick disks and halos can be seen in yellow in the external parts. Each box has a size of 100 kpc.}
\label{Fig_edgeonview_faceonview} 
\end{figure*}

Figure \ref{Fig_edgeonview_faceonview} shows an atlas of face-on
and edge-on views of the 12 remnant galaxies part of the core study,
$\sim 8$ Gyr after the first passage and $\sim 6.3$ Gyr after fusion.
The remnants are all spiral galaxies with several structures such as
bulges, large disks, bars, or rings. It emphasizes that no specific needs are required to rebuild disks after realistic gas-rich mergers expected to occur at moderate to high-redshift.  It has been often proposed that disk rebuilding only occurs in the presence of extreme or peculiar feedback conditions in the central regions, for, e.g., preserving the gas before fusion. We verified that decreasing our feedback to the median value adopted by \citet{Cox2006} would not change drastically our conclusions : for realistic values of the gas fractions in progenitors taken before the first passage at z $\sim$ 1.5, major mergers produce only spiral galaxies.  They were found to have stellar
masses (within r = 15 kpc) in the range $3.99\times10^{10} M_{\sun} < M_{stellar} <  5.47\times10^{10}  M_{\sun}$, which is in agreement
with the mass range of the observational samples (see \S2). Resulting
$B/T$ are given in Table~\ref{Table_fit} (see the 12 first entries and also compare with the next set of 12 similar simulations with 2 M particles) showing that
all remnants in the core study have B/T $<$ 0.45. Since in the two components decomposition, bar is assumed to be part of the central region (see \S2.3), these B/T values correspond to upper limits of the bulge-to-total fraction.

\subsection{Properties of bulges \& bars in the core study}
\label{bulge_bar}
\begin{figure}
\centering
\includegraphics[width=0.5\textwidth]{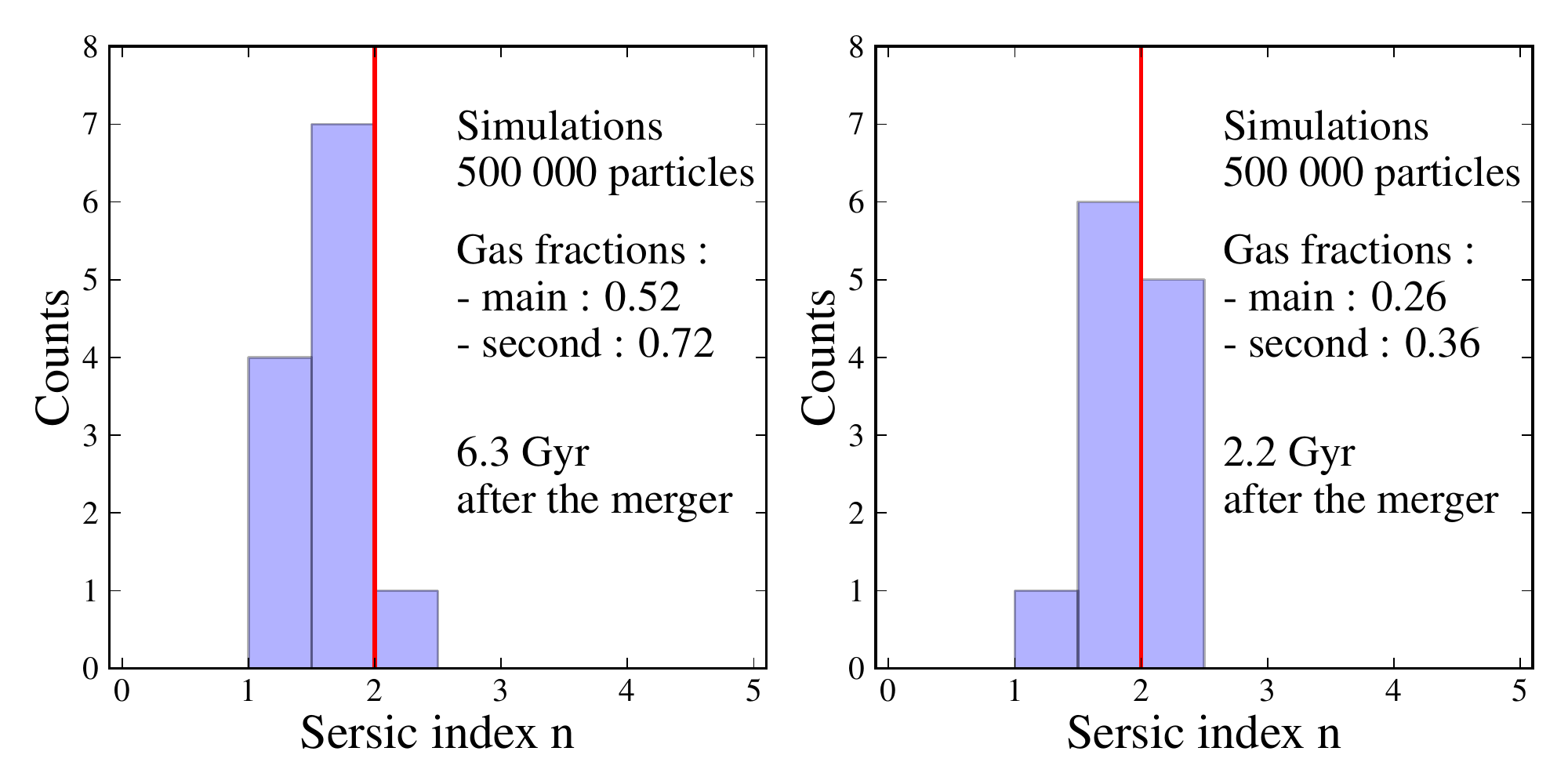} 
  \caption{Sersic index distribution for the core study assuming a fusion occurring 6.3 Gyr ago ({\it left}) and 2.2 Gyr ago ({\it right}). Corresponding gas fractions are indicated (see also text). The cut at n = 2 is shown as a red line (in the online version).}
\label{Fig_principale_histo2}
\end{figure}

Figure \ref{Fig_principale_histo2} shows the distribution of the Sersic indices
($n$), which are almost all characteristic of pseudo-bulges. We attribute the fact that
we have very few classical bulges in the remnants
to the large gas fractions adopted in the core study, which
are representative of conditions in progenitors at z $\sim$ 1.5 (see
\S3.1). Indeed, high gas fractions dump the violent relaxation during the fusion. Later on, the gas
may accumulate in the center and can form new stars in the surroundings of the newly formed bulge
 \citep{Hopkins2009,Hopkins2010}. This phenomenon of gas feeding the center may resemble a secular process, while instead, it is actually a direct consequence of the merger.

Does it mean that mergers are only producing pseudo-bulges? In Section~\ref{variations_gas_mass} we will show that higher gas mass fractions expected for mergers occurring at higher redshifts than those in the core study, would unavoidably lead to more numerous pseudo-bulges. However some mergers may have occurred more recently, and from progenitors having smaller gas fractions.    
To test this, we simulate similar mergers than those of the core study but with main and secondary
progenitors having gas fractions of only 26\% and 36\%, respectively. The latter corresponds to gas fractions representative of $z\sim0.5$ galaxies
\citep{Rodrigues2012}, and the morphological
analysis was repeated only 2.2 Gyr after fusion (or 5.2 Gyr
after the beginning of the simulation) to keep these mergers representative of the cosmic epoch - gas fraction relationship. 
Figure \ref{Fig_principale_histo2} evidences significant larger values for the Sersic indices, with 40\% of classical bulges instead of a few percents. We attribute such a large change to the combination of two factors, one being the decreased gas fraction, the other because later-on gas feeding the central regions would have less time to proceed. However the observed evolution of the merger rate highly favors merger events occurring between high-redshift gas-rich galaxies. Then, the main result of the core study stands : major mergers occurred mostly at high redshifts and should mostly lead to form pseudo-bulges.

\begin{figure}
\centering
\includegraphics[width=0.5\textwidth]{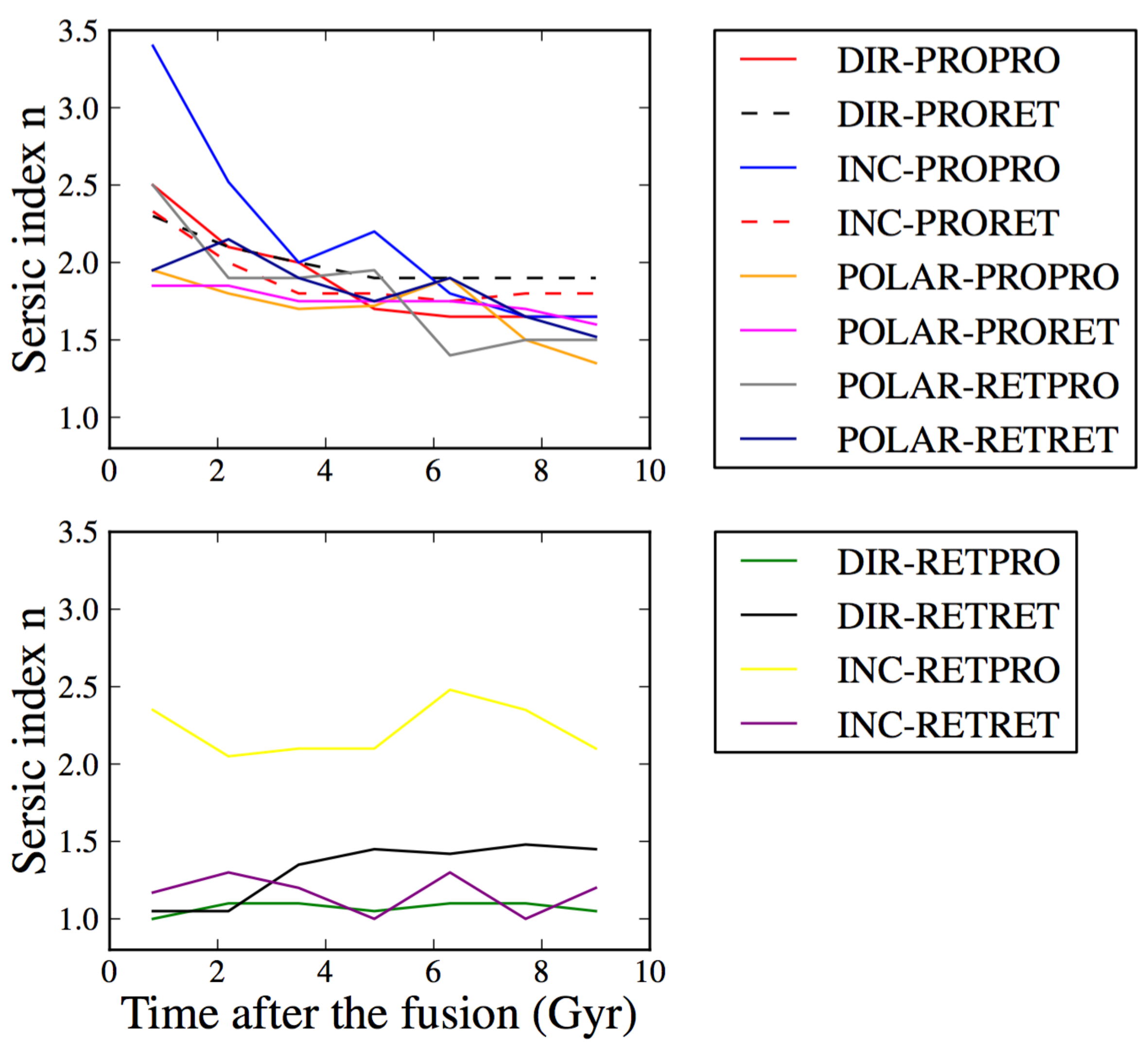}
  \caption{Time evolution of the Sersic indices for the core study for the different orbits (see the color version of this figure in the online journal to distinguish each orbit).}
\label{Fig_time}
	\end{figure}

\begin{figure}
\centering
\includegraphics[width=0.55\textwidth]{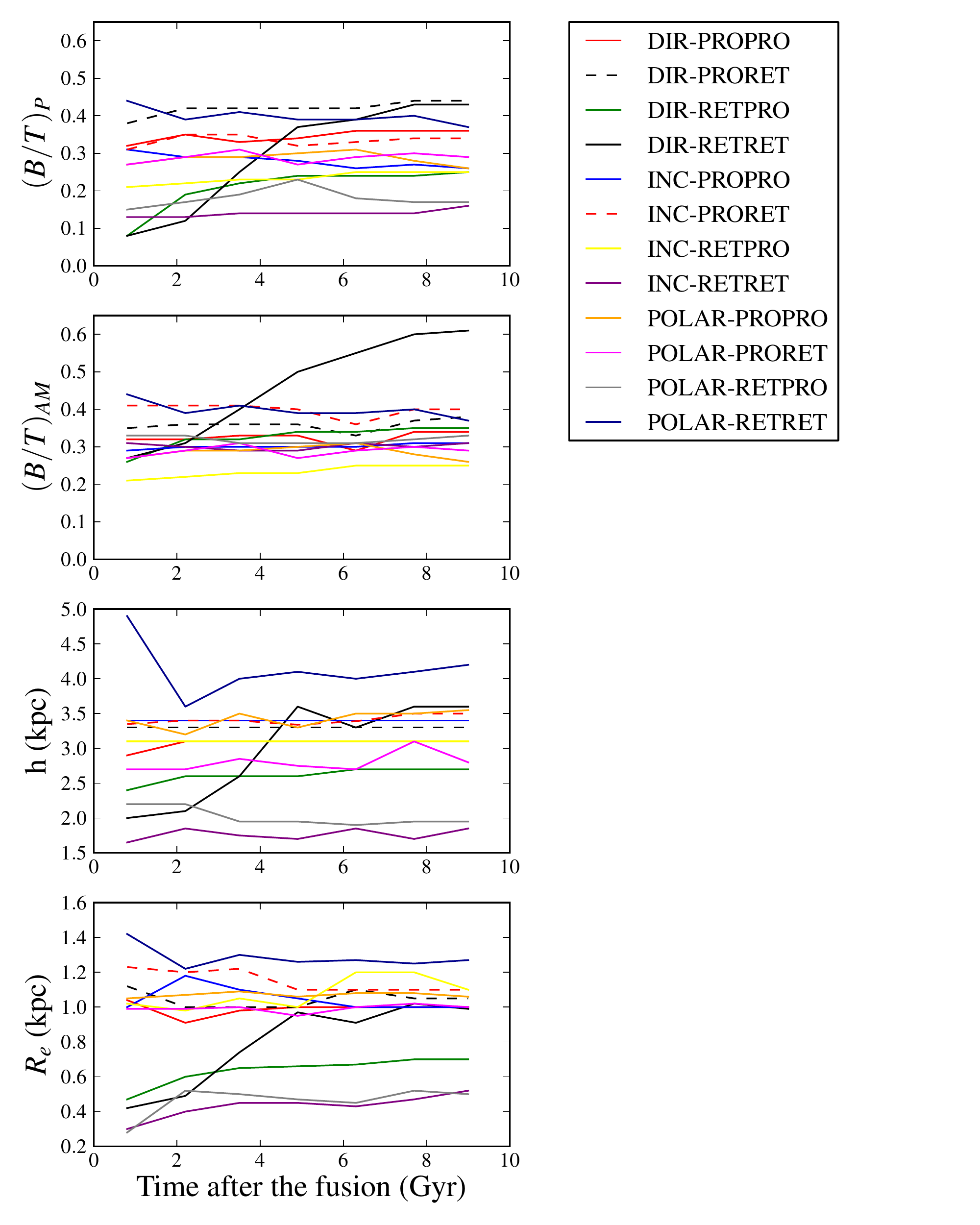}
  \caption{Time evolution of the parameters $(B/T)_{P}$, $(B/T)_{AM}$, $h$ and $R_{e}$ for the core study as a function of the different orbits (see the color version of this figure in the online journal to distinguish each orbit).}
\label{Fig_time2}
	\end{figure}

We repeated the morphological decomposition at 0.8, 2.2, 3.5, 4.9,
6.3, 7.7, and 9.0 Gyr after fusion. Figure \ref{Fig_time} shows the
resulting time evolution of the Sersic index for the core study. 
Every simulation presents a bar at some time in the simulation, except 
the simulation with the orbit DIR-PROPRO. 
For eight simulations (see Figure \ref{Fig_time}, top panel), the Sersic index decreases  
with time, in contrast with most other parameters that remained unchanged (see Figure \ref{Fig_time2}). 
This behavior suggests that gas is accumulating towards the central regions often driven by bars, resulting in the formation of an increasingly disk-like mass profile characterizing pseudo-bulges. 
Figure \ref{Fig_time} indicates that even if a classical bulge ($n$ $>$ 2) is formed $\sim$ 1 Gyr after the fusion, a pseudo-bulge component is progressively superimposed, leading to smaller Sersic index  (see also \citealt{Athanassoula2016}). Four orbits (see Figure \ref{Fig_time}, bottom panel) however show 
no evolution with time of the Sersic index that remains constant, including the INC-RETPRO orbit, which kept a permanent classical bulge. 
The orbital parameters of this specific simulation influence the gas distribution 
and prevent significant amount of gas to be brought into the center. 
We conclude that gas-rich major mergers can form classical and pseudo-bulges, with an increasing contribution of the later with time, both coexisting as it has been suggested for the Milky Way bulge \citep{Babusiaux2016}.

\begin{figure}
\centering
\includegraphics[width=0.5\textwidth]{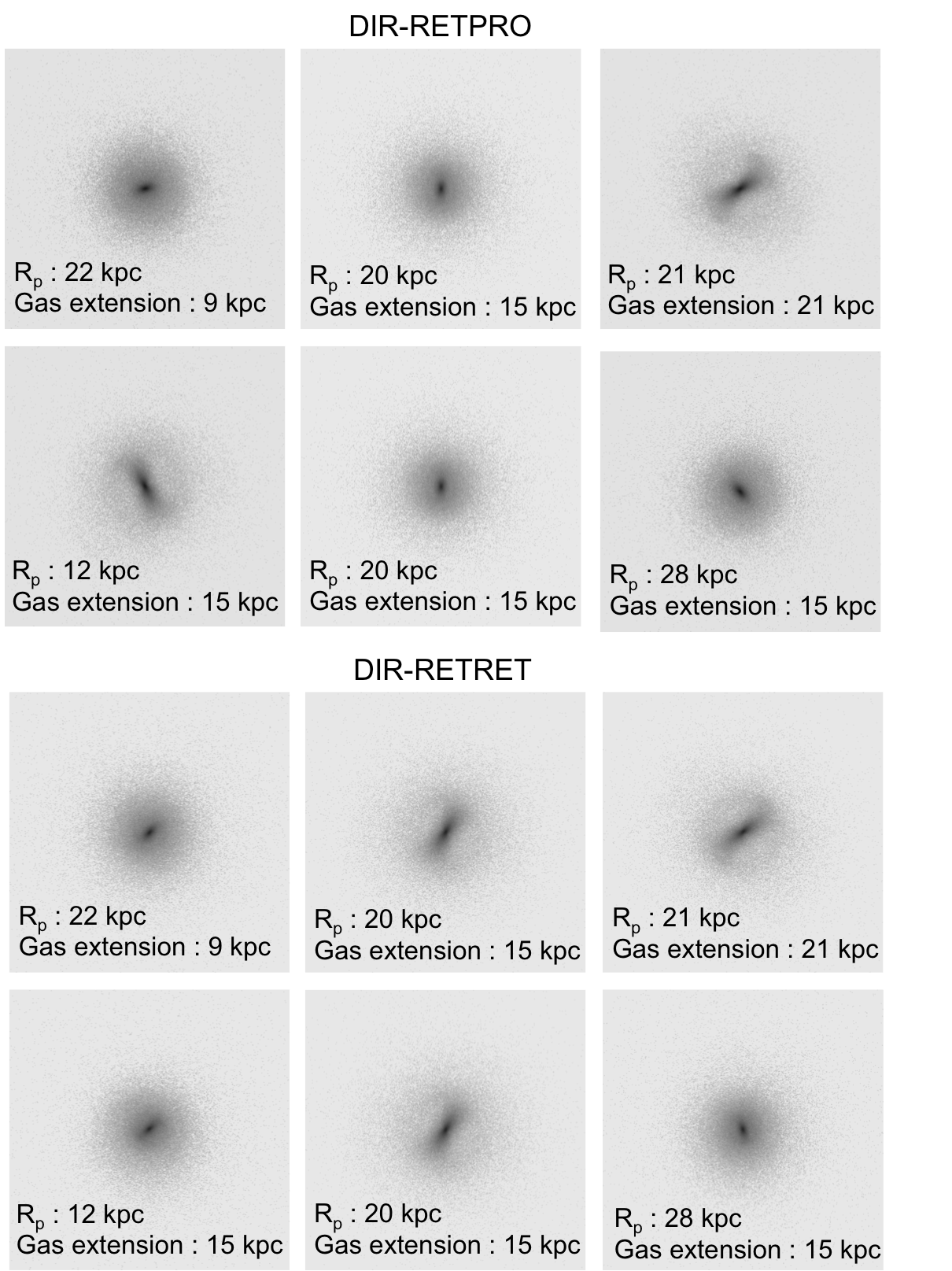} 
  \caption{2D images of the two simulations DIR-RETPRO and DIR-RETRET favorable to form bars. For each panel, on the top we have changed the gas extension, while on the bottom we have changed the pericenter ($R_p$). }
\label{Fig_gasperi}
\end{figure}

Almost two thirds of local spiral galaxies have bars in their central
parts \citep{Eskridge2000, Knapen2000, Laurikainen2004, Menendez2007,
  Barazza2008, Aguerri2009, Gadotti2009}, which result from disk
instabilities \citep{Binney2008}.
\noindent During a major merger, 
it is also common to form bar-like structures at the beginning of the merger 
during the first passage resulting from resonances between the main progenitor 
and the orbit \citep{Hopkins2009, Athanassoula2016}. At these stages, bars may have a high angular momentum and rotate
relatively fast. If the resonance is strong the bar can survive after the fusion and 
bring gas into the central parts of the remnant. It can then evolve
and change in orientation, spin, shape, and length. Bars 
have Sersic index between 0.5 and 1, as expected since they are
density fluctuations of the disk with which they share similar
properties (see \citealt{Gadotti2011}).

In the core study, we obtained $9/12$ orbits forming bars, whose 2 with large
bars (size $\ge 4$ kpc for DIR-RETPRO and DIR-RETRET), 2 other with medium size
bars ($2 <$ size $< 4$ kpc for INC-RETPRO and INC-RETRET). The last 5 bars have size $\le 1.1$ kpc. For the two simulations with large bars, we explored the impact of the pericenter (12, 20 and 28
kpc) and gas extension (9, 15 and 21 kpc) on the bar formation. Figure
\ref{Fig_gasperi} shows that for both orbits, a stronger bar is obtained
when the extension of gas is equal or less than the pericenter value,
probably because of more favorable resonances.
Note that these configurations imply that when the secondary
progenitor is at pericenter, both progenitor gas disks are in contact.
More tests on different orbits would be necessary to conclude on the
resonance between the gas extension and pericenter.

\subsection{Influence of gas fraction \& mass ratio}
\label{variations_gas_mass}

\begin{figure}
\centering
\includegraphics[width=0.55\textwidth]{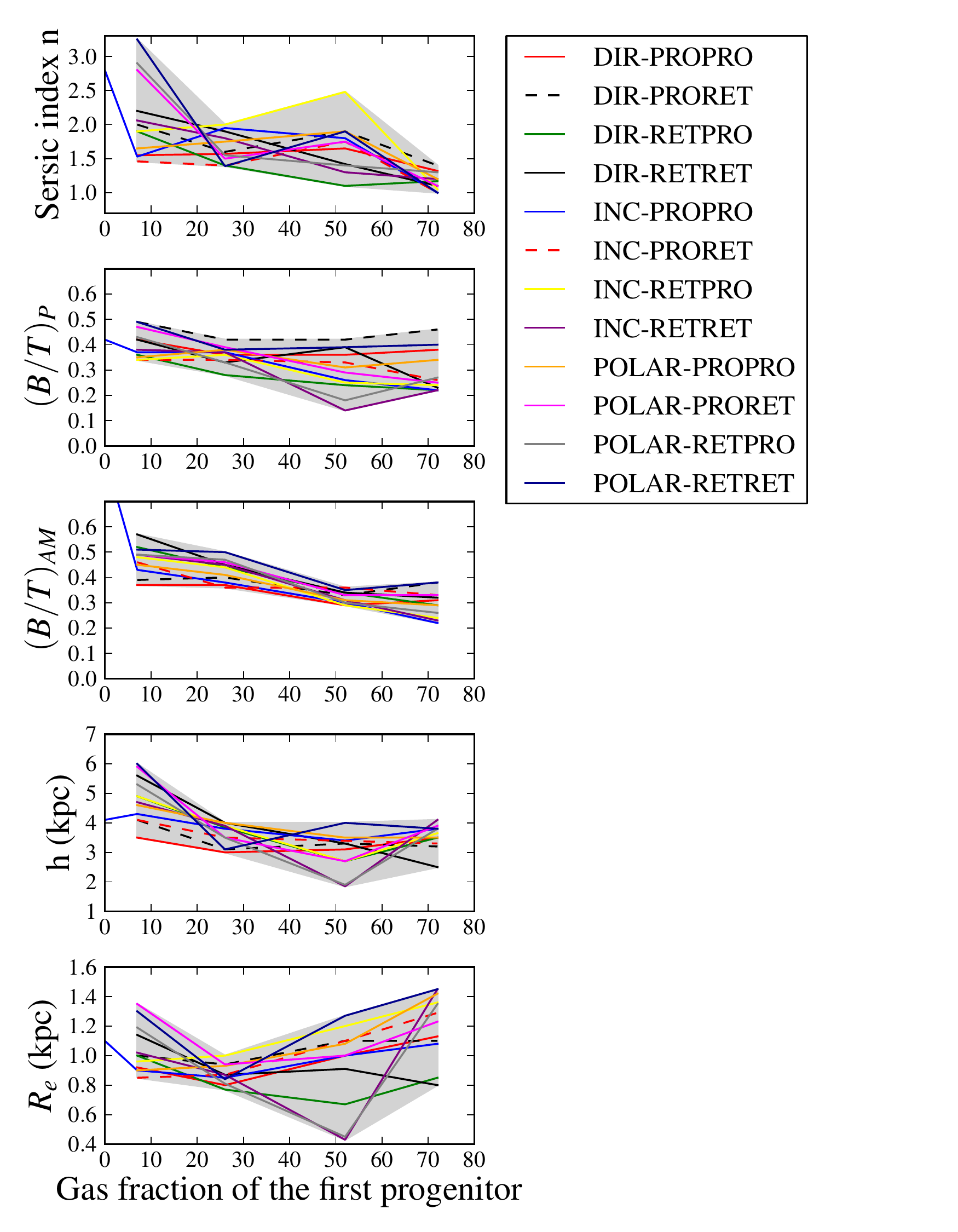}
  \caption{{\it From top to bottom} : Sersic index, photometric B/T, angular momentum B/T and size of disks (h) as a function of the main progenitor gas fraction (see text). We show the dependance of these parameters for the twelve simulations with 500 000 particles (see the color version of this figure in the online journal to distinguish each orbit).}
\label{Fig_gasfraction}
\end{figure}

We tested three additional gas fractions for the main and secondary
progenitors : $7\%$-$10\%$, $26\%$-$36\%$, and $72\%$-$92\%$,
respectively. Figure \ref{Fig_gasfraction} shows the resulting Sersic
indices, 1D-fitting morphological $B/T$ ratios, alternative $B/T$
ratios measured using an angular momentum decomposition following
\citet{Hammer2010}, and disks scale lengths $h$ of the thin disks as a
function of gas fraction. The $(B/T)_{AM}$ calculated from the angular momentum (i.e., all particles that are lying in the bulge radius and showing no preferential angular momentum) is determined with a much higher accuracy that the photometric $(B/T)_P$ and corresponds to a more physical parameter to characterize merger remnants. This is why it is interesting to see that almost all models show very similar $(B/T)_{AM}$ ratio (see also Figure \ref{Fig_time2}) and that this ratio is decreasing with increasing gas fraction. However $(B/T)_{AM}$ cannot be properly estimated in most observations, so in this paper it serves us as a guidance to control possible effects linked to the $(B/T)_P$ estimates that show much larger scatter (see a comparison between the two ratios in Appendix~\ref{appen:bt}).

Figure \ref{Fig_gasfraction} shows that in general all parameters are decreasing with increasing gas fraction, but the bulge effective radius that shows a decrease and then an increase for gas fraction larger than 30\%. We suspect that the later behavior is associated to the large bars, whose formations are favored by large gas fractions. The same mechanism probably explains why the Sersic index of all remnants is close to 1 for gas fractions $72\%$-$92\%$.

\begin{figure}
\centering
\includegraphics[width=0.55\textwidth]{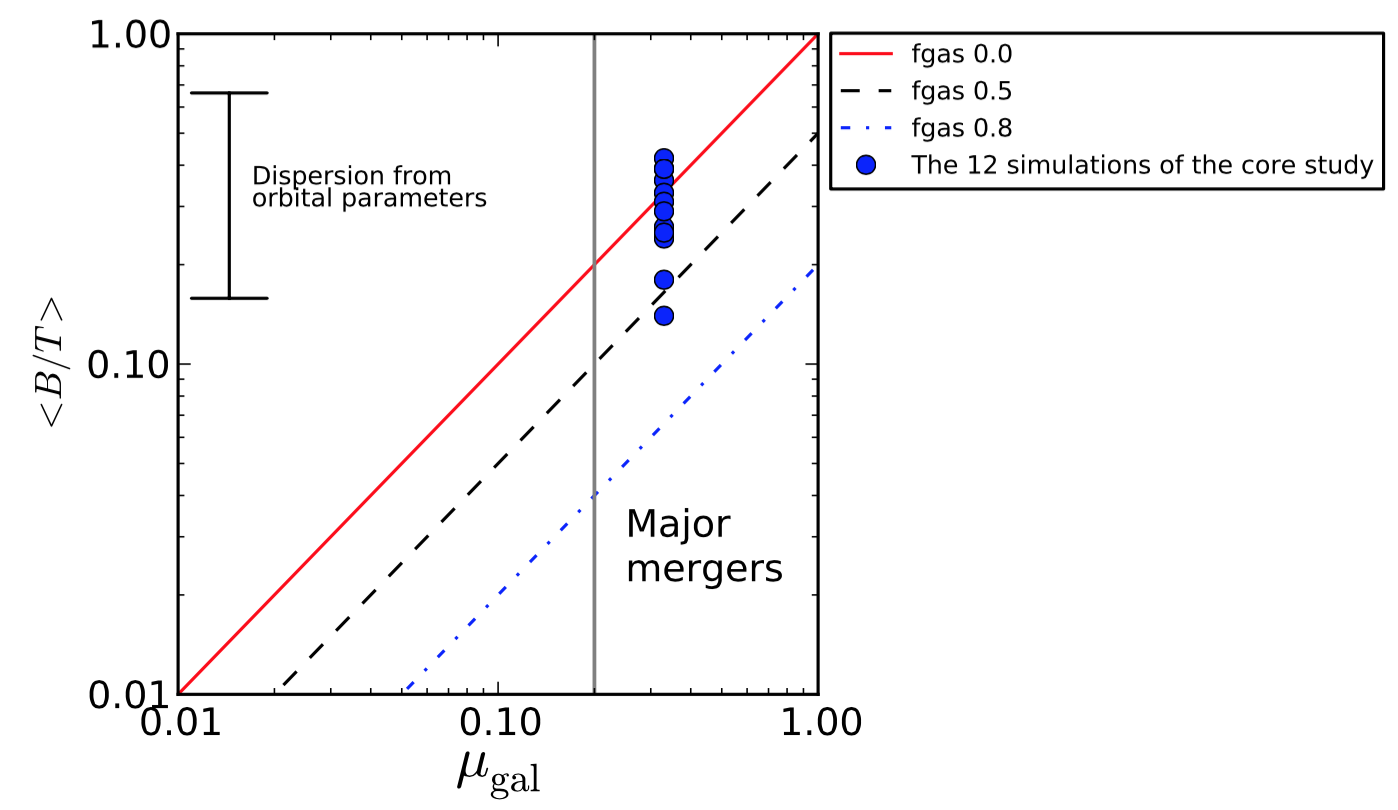} 
  \caption{Mean values of $(B/T)_P$ from the 12 orbits of the core study for 4 different gas fraction compared to the dispersion given by the relation $B/T \sim \mu_{gal}(1-f_{gas})$ from \citet{Hopkins2010} where $\mu_{gal}$ is the mass ratio between the two progenitors. The vertical line corresponds to a mass ratio 0.2, which gives a limit between major mergers and minor mergers.}
\label{Fig_hopkins}
\end{figure}

Figure \ref{Fig_hopkins} shows the resulting morphological $B/T$
superimposed on the results from the study of \citet{Hopkins2010}. 
The mean $(B/T)$ over all orbits is 0.30 with a standard deviation of
0.08. The scatter in $B/T$ associated to a change of orbital parameters
is found to be similar to the scatter associated to the change of gas fraction
in the progenitors as determined by \citet{Hopkins2010}. We however notice that our B/T estimates are systematically larger than those theoretically calculated by \citet{Hopkins2010} since the later correspond to the sole classical bulge component conversely to our estimates that include contribution from bars.


We found that the higher the gas fraction, the larger the
 bars are.  Table \ref{Table_fit} shows that we obtain no large bars 
(size $\ge 4$ kpc) with gas fractions of
$7\%-10\%$ and of $26\%-36\%$, while 2 large bars are obtained for gas fraction of
$52\%-72\%$ and 5 for gas fractions of $72\%-92\%$. 

For verifying the ability of SPH merger models to form bulge-dominated galaxies, we also conducted simulations with gas fraction as
small as $0.2\%$ in the main progenitor and $0.6\%$ for the secondary,
and for a mass ratio of 3:1 and 1:1. In both cases we obtain a high
Sersic index ($n > 2.5$), while a bar form in the center. A thick disk
is formed in the first case, while an elliptical galaxy is obtained in
the second.

Finally, we also tested two other different mass ratios for three
particular orbits, i.e., 1.5:1 and 4.5:1. We note that for the orbit favorable to bar formation (DIR-RETPRO), the bar is larger when the mass ratio is 4.5:1 than when the mass ratio is 1.5:1.

\section{DISCUSSION \& CONCLUSION}
\indent We have presented a systematic study of gas-rich major merger remnants to verify whether or not there is a tension between the low fraction of nearby galaxies with classical bulges and the hierarchical scenario of galaxy formation. The main result of this study is that mergers of galaxies conducted in realistic cosmological conditions are producing mostly pseudo-bulges and only few classical ones. Indeed most mergers are expected to occur in the distant Universe where galaxies are gas-rich, and our study shows that the residual gas after the fusion gradually falls into the center forming bars and pseudo-bulges for most of the orbital parameters. Forming a significant fraction of classical bulges requires mergers to occur in the nearby Universe with fusion times a few billion years ago, but these events are unlikely to be dominant in the hierarchical scenario \citep{Lefevre2000, Rawat2008}. Formation of bulge-dominated galaxies with Milky Way masses requires event more stringent conditions, i.e., extremely small gas fractions and preferentially 1:1 mergers.\\
\indent We argue that our study is rather representative of expectations from cosmological conditions, because (1) progenitor gas fractions are adapted from their epoch of involvement into a merger; (2) the orbital parameters are representative of cosmological simulations presenting the whole range of angles expected for parabolic encountering.  Our choice of mass ratio to 3:1 (from \citealt{Hammer2009}) of 16 kpc pericenter and of high feedback is not affecting our main result since variations of these parameters do not change our conclusions. Perhaps a significant progress in sampling properties of mergers remnants would be to allow variance in the orientation of the secondary (assumed to be polar), which would let us with 24 additional simulations in the core study. Otherwise we may conclude that our approach is useful when compared to cosmological simulations since it allows to characterize remnants with a large number of particles and with a rather realistic treatment of the gas hydrodynamics.\\
\indent All remnants of our core study have been analyzed using a two-component (disk+bulge) method for fitting their surface mass-density profiles, in a very similar way to what has been done for observed galaxies, and with similar limitations on spatial resolutions. It results that all remnants of our core study are spiral galaxies, with their centers dominated by pseudo-bulges and bars, instead of classical bulges. This leads to another important result of this study, since no specific recipe on feedback has been required, and we have verified that reducing feedback to the mean value of \citet{Cox2006} does not significantly affect it. A significant improvement of our modeling and comparison with observations may be to include realistic dust and stellar populations, even if we have shown that dust effects could not have strongly biased our statistics (see Section \ref{section:delgado}).\\
\indent Further tests could be done to characterize even more realistic hydrodynamics conditions with, e.g., an hydrodynamical solver such as GIZMO \citep{Hopkins2015}. Higher number of particles would be also useful to compare the properties of the remnants with those of Local galaxies such those studied by  \citet{Kormendy2010}.\\
\indent This study is also bringing a data basis of how should be realistic merger remnants for, e.g., verify whether their properties resemble that of today spiral galaxies. For example, one may study the double disk properties (as done by \citealt{Peschken2017}) to verify if the observed fractions of Type I, II and III are consistent with our results. Figures \ref{Fig_time2} and \ref{Fig_gasfraction} evidence that orbits provide a large variance in B/T, disk scale-length, and bulge effective radius, which could represent qualitatively well most of the present-day spirals : in fact we reproduce most of their structures, such as thin and thick disks, halo, bulges, bars and rings.\\
\indent Finally we conclude that the dichotomous picture for which pseudo-bulges form secularly while classical bulges form in mergers, has been perhaps pushed too much, and we hope this study may help to clarify that at least pseudo-bulge may form in another way. An important follow up of this paper will be to compare the overall properties (including luminosity profiles and kinematics) of galaxies with the results of a grid of realistic merger remnants, at different epochs.
\section*{Acknowledgments}
We acknowledge the financial support of T.S. PhD grant from PSL (Paris Sciences et Lettres - France).
This work was granted access to the HPC resources of MesoPSL financed
by the Region Ile de France and the project Equip@Meso (reference
ANR-10-EQPX-29-01) of the program 'Investissements d'Avenir' supervised
by the Agence Nationale pour la Recherche.
We wish to thank Jianling Wang for helpful comments, discussions and carefully reading this manuscript. We wish also to thank Karen Disseau for her help in programming and discussions on galaxies morphology. We are grateful to the referee for useful comments and suggestions, which have greatly improve the content of this manuscript.



\bibliographystyle{mnras}





\appendix

\section{Methodology of the morphological study on the Delgado-Serrano sample.}

\label{appen:degado}

\begin{figure*}
\centering
\includegraphics[width=1.0\textwidth]{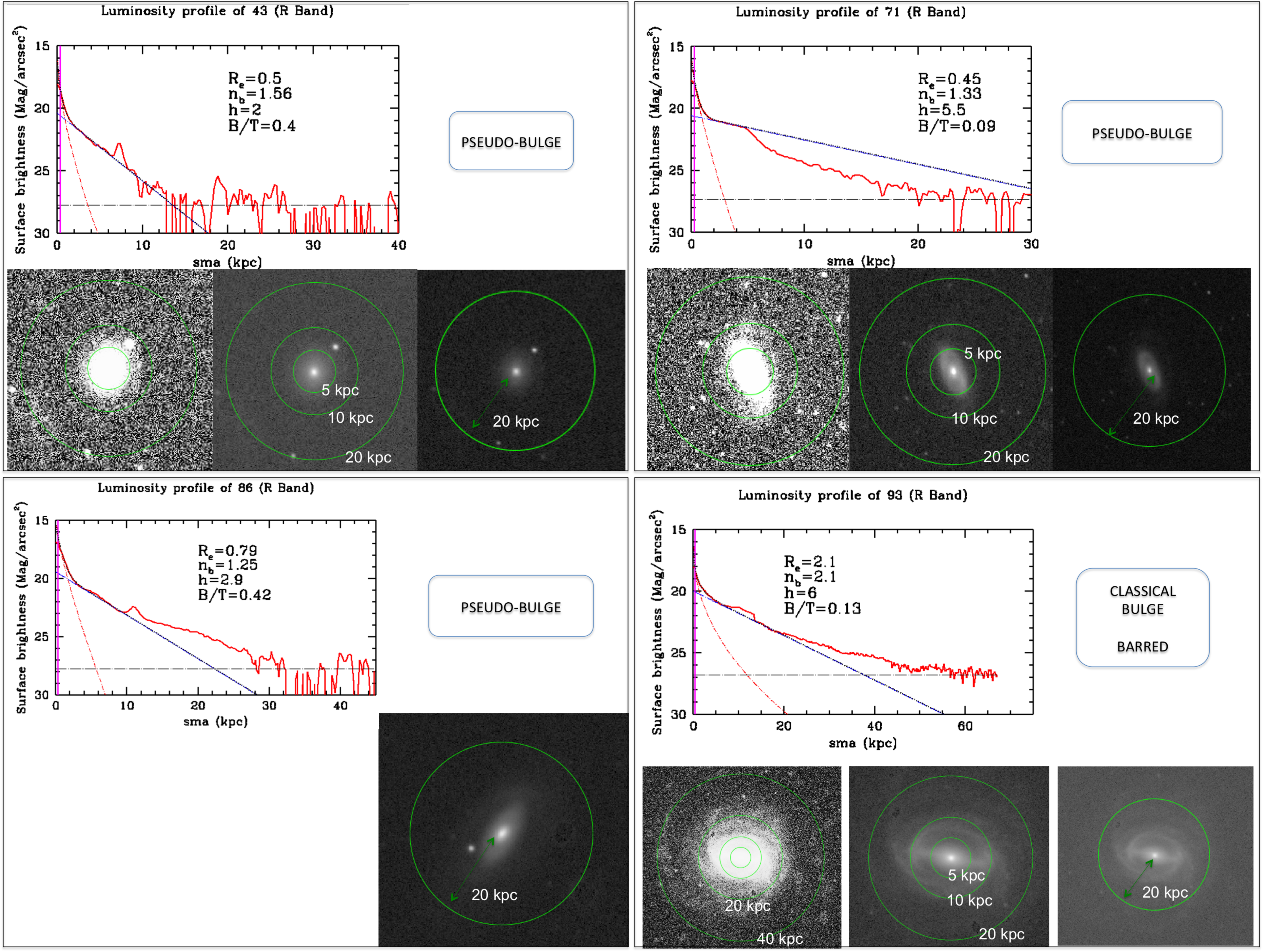}
  \caption{Morphological analysis for four galaxies in the Delgado sample. The thick red line corresponds to the surface brightness of the object (a color version of this figure is available in the online journal), the dash red line is the fitted Sersic component, the dash blue line is the fitted exponential disk. The magenta line represents the HWHM of the PSF ($\sim 1.4$ arcsec). The main parameters associated to the fit are given on the plot, and the bulge classification is provided inside the blue box on the right. Bottom panels show the galaxies with different contrasts providing also concentric circles to compare with the fit of the profile.}
\label{Fig_fit_delgado}
	\end{figure*}

For each galaxy of the Delgado-Serrano et al. (2010) sample, we have revisited the estimation of the B/T. To fit the profile of the corresponding 66 objects, we have first recovered the profile of each galaxy using the Ellipse task within IRAF. The recovered isophotal profile was then fitted using an interactive program developed by our team that is based on a two components fit (bulge+disk). The central region of each profile was not used because the flattening in the central region is only due to the limited spatial resolution of the observations (seeing). All the fits were performed by TS. Each fit was validated independently by FH and MP, a final sampling verification was performed by HF. \\ 
Figure \ref{Fig_fit_delgado} presents four examples of disk+bulge decompositions in 1D (see captions for details). Table \ref{table_delgado} displays the parameters of the Delgado-Serrano et al. (2010) galaxies.
  To have an estimation of how the flux profile in the central parts of the galaxies is affected by the spatial resolution,
  we degraded HST images of a very nearby galaxy (NGC 3982) down to the SDSS PSF. We verified that for bulges with effective radius smaller than 3 to 7 times the HWHM, the peak at the center could be underestimated by up to one magnitude. 
  We circumvent this problem by avoiding any fit within the HWHM (see the magenta line in Figure \ref{Fig_fit_delgado}), and after further tests in fitting bulges we estimate to $\pm$ 0.2 the error on the Sersic n index.
  
	
\begin{landscape}
\begin{table}
\begin{tabular}{lcccccccccc}
\hline
\hline
Num  & Name &RA (deg) &DEC (deg) &z   &n &$R_e$ &h  &$(B/T)_P$ &Bulge  type &Bar?     \\
   & & & & &$\pm 0.20$ &$\pm 0.10$   &$\pm 0.30$ &$\pm 0.10$  & &       \\          
\hline
     4 &$J094847.11+010311.5\_59$ &147.1958 &  1.0531  &0.0273 & -- & -- & 1.03 & -- &Bulgeless & \\
   5 &$J094913.51+011039.8\_61$ &147.3042 &  1.1778  &0.0251  & 1.35 & 0.35 & 1.70 & 0.10 &Pseudo & \\
       8 &$J103135.16+002831.3\_434$ &157.8958 & -0.4753  &0.0287 & 1.70 & 0.90 & 4.60 & 0.08  &Pseudo &Y\\
       10  &$J103534.47-002116.2\_449$  &158.9139 &-0.3155  &0.0293  &1.05 &0.45 &1.50 &0.28 &Pseudo &\\
         11 &$J103614.28-000932.4\_454$ &159.0583 & -0.1589  &0.0277 & 1.20 & 0.28 & 2.30 & 0.06 &Pseudo &\\
      12 &$J103657.37+001347.0\_457$ &159.2375 & -0.2297  &0.0292  & 1.25 & 0.45 & 4.80 & 0.05 &Pseudo  &\\
      16 &$J110839.61+001703.5\_680$ &167.1825 &0.3142 &0.0248  &1.36 &0.76 &2.20 &0.35 &Pseudo &\\
  17 &$J110840.90+002330.2\_682$ &167.1708 & -0.3917  &0.0253   & -- & -- & 1.80 & -- &Bulgeless &Y\\
  18 &$J110914.81-005540.5\_686$ &167.3322 &-0.9448 &0.0288 &2.00 &1.75 &7.00 &0.72 &Classical  &\\
  20 &$J111202.58-001029.3\_715$ &168.0125 & -0.1750  &0.0249 & 1.40 & 0.52 & 1.61 & 0.29 &Pseudo &\\
    21 &$J111339.88+004915.5\_730$ &168.4167 & -0.8208  &0.0286   & 1.50 & 0.60 & 1.90 & 0.15 &Pseudo &Y\\
  23 &$J111559.49-002059.0\_751$ &169.0000 & -0.3497  &0.0260  & 1.40 & 0.40 & 2.15 & 0.06 &Pseudo &Y\\
24 &$J111609.33+003424.7\_754$ &169.0375 & -0.5736  &0.0290 & 1.00 & 0.37 & 3.80 & 0.04 &Pseudo & \\
   27 &$J111849.55+003709.3\_772$ &169.7042 & -0.6192  &0.0254  & -- & -- & 1.60 & -- &Bulgeless  & \\
   31 &$J112408.63-010927.8\_843$ &171.0375 & -1.1578  &0.0293 & 1.50 & 0.70 & 6.08 & 0.23 &Pseudo & \\
     32 &$J112409.18+004202.0\_844$ &171.0375 & -0.7006  &0.0260 & 1.50 & 0.80 & 6.70 & 0.03 &Pseudo &Y\\
    33 &$J112418.64+003837.4\_846$ &171.0792 & -0.6436  &0.0264   & 1.85 & 1.06 & 2.80 & 0.13  &Pseudo &Y\\
    34 &$J112437.05-005930.7\_848$ &171.2241 &-0.9447 &0.0256  &2.00 &2.25 &6.70 &0.62 &Classical &\\
    35 &$J112535.07-004605.6\_854$  &171.389  &-0.7341   &0.0253    &2.75  &7.00  &21.00  &0.78  &Classical &\\    
  37 &$J112738.15+003950.9\_864$ &171.9083 & -0.6642  &0.0291   & 1.77 & 0.80 & 2.09 & 0.21 &Pseudo & \\
  39 &$J113116.65+001323.3\_882$ &172.8208 & -0.2231  &0.0297 & 1.32 & 0.50 & 1.80 & 0.20 &Pseudo &  \\
  40 &$J113209.23-005633.3\_886$ &173.0375 & -0.9425  &0.0262   & 1.47 & 0.70 & 2.95 & 0.30 &Pseudo &Y\\
  42 &$J113420.50+001856.4\_896$ &173.6190 &0.3143 &0.0289 &2.00 &1.70 &8.00 &0.62 &Classical &\\
  43 &$J113438.36+001119.1\_898$ &173.6583 & -0.1886  &0.0289 & 1.56 & 0.50 & 2.00 & 0.40  &Pseudo & \\
  44 &$J113439.15+000729.1\_899$ &173.6625 & -0.1247  &0.0288  & 1.75 & 0.74 & 3.10 & 0.02 & Pseudo  & \\
  46 &$J113523.27+000525.9\_902$ &173.7852 &0.1051 &0.0292 &2.10 &1.85 &6.00 &0.72 &Classical &\\
  47 &$J113833.27-011104.1\_914$ &174.6375 & -1.1844  &0.0208 & 1.65 & 0.45 & 2.97 & 0.03  & Pseudo & \\
  48 &$J114006.20-005405.0\_921$ &175.0250 & -0.9014  &0.0287 & 1.30 & 0.62 & 6.00 & 0.10 &Pseudo & \\
  50 &$J114453.04+005622.6\_935$ &176.2208 & -0.9397  &0.0280  & 1.20 & 0.44 & 2.00 & 0.24 &Pseudo & \\
   52 &$J120023.91+002926.4\_1007$ &180.1000 & -0.4906  &0.0257  & 1.70 & 0.90 & 2.85 & 0.11 &Pseudo & \\
  53 &$J120127.92-004306.1\_1014$ &180.3667 & -0.7183  &0.0207  & 1.84 & 0.90 & 4.76 & 0.07 &Pseudo &Y\\
      54 &$J121411.31-004953.6\_1086$ &183.5458 & -0.8317  &0.0249  & 1.00 & 2.30 & 3.72 & 0.09 &Pseudo &Y\\
       56 &$J121752.89-003926.0\_1102$ &184.4708 & -0.6572  &0.0296 & -- & -- & 2.50 & -- &Bulgeless & \\ 
\hline
\hline
\end{tabular}
\label{table_delgado}
\caption{Results of 1D fitting (1/2) for the 66 spiral and S0 galaxies of the \citet{Delgado2010} sample.}
\end{table}
\end{landscape}

\begin{landscape}
\begin{table}
\begin{tabular}{lcccccccccc}
\hline
\hline
Num  & Name &RA (deg) &DEC (deg) &z    &n &$R_e$  &h  &$(B/T)_P$ &Bulge type  &Bar?    \\ 
   & & & & &$\pm 0.20$ &$\pm 0.10$   &$\pm 0.30$ &$\pm 0.10$  & &       \\        
\hline
\hline
58 &$J122337.48-002821.3\_1124$ &185.9042 & -0.4725  &0.0256  & 1.00 & 0.2 &2.02  &0.00 &Pseudo &Y\\
  60 &$J122411.84+011246.8\_1127$ &186.0500 &  1.2131  &0.0266 & 1.60 & 0.40 & 3.80 & 0.01 &Pseudo  & \\
  64 &$J123846.47+001921.0\_1181$ &189.6958 & -0.3225  &0.0228  & 1.90 & 1.33 & 3.50 & 0.13 &Pseudo &Y\\
 71 &$J132903.23-000237.0\_1488$ &202.2625 & -0.0436  &0.0214  & 1.33 & 0.45 & 5.50 & 0.09 &Pseudo & \\
  72 &$J134822.64-004601.1\_1586$ &207.0958 & -0.7669  &0.0261 & 1.25 & 0.30 & 1.20 & 0.04 &Pseudo  &\\
  73 &$J135102.22-000915.1\_1594$ &207.7583 & -0.1542  &0.0235  & 1.92 & 1.00 & 4.70 & 0.13 &Pseudo &Y\\
 75 &$J135342.79+000339.1\_1611$ &208.3883 &0.1033 &0.0299 &2.15 &3.00 &12.00 &0.82 &Classical &\\
  77 &$J135807.05-002332.9\_1624$ &209.5292 & -0.3925  &0.0296  & 1.85 & 1.50 & 4.10 & 0.13 &Pseudo & \\
  78 &$J140013.27-005745.8\_1634$ &210.0542 & -0.9628  &0.0249 & 1.30 & 0.57 & 1.95 & 0.32  &Pseudo & \\
  80 &$J140320.74-003259.7\_1643$ &210.8375 & -0.5581  &0.0247  & 1.10 & 0.37 &11.00 & 0.01 &Pseudo  &  \\
  82 &$J140451.73-003829.7\_1651$ &211.2167 & -0.6417  &0.0244  & 1.57 & 0.80 & 4.00 & 0.16 &Pseudo &Y\\
  83 &$J140452.62-003640.5\_1652$ &211.2208 & -0.6111  &0.0248   & 1.70 & 0.74 & 5.50 & 0.06 &Pseudo  &\\
 84 &$J140601.42-001837.9\_1660$ &211.5042 & -0.3106  &0.0246 & 1.60 & 1.70 & 7.00 & 0.13 &Pseudo &Y\\
 85 &$J140831.60-000737.3\_1675$ &212.1333 & -0.1269  &0.0251  & 1.00 & 0.40 & 2.25 & 0.00 &Pseudo &Y\\
  86 &$J141026.83-004956.5\_1682$ &212.6125 & -0.8322  &0.0255   & 1.25 & 0.79 & 2.90 & 0.42 &Pseudo & \\
  92 &$J141814.91+005327.9\_1748$ &214.5625 & -0.8911  &0.0260  & 1.40 & 0.55 & 4.10 & 0.05 &Pseudo & \\
  93 &$J142720.36+010133.0\_1798$ &216.8333 &  1.0258  &0.0257 & 2.10 & 2.10 & 6.00 & 0.13  &Classical &Y\\
  94 &$J142804.53+010022.6\_1805$ &217.0208 &  1.0064  &0.0256  & -- & -- & 1.80 & -- &Bulgeless &\\
  96 &$J143101.75+011434.2\_1830$ &217.7583 &  1.2428  &0.0262  & 1.55 & 0.55 & 2.20 & 0.11  &Pseudo & \\
  97 &$J143124.59+011403.7\_1832$ &217.8059  &1.1587  &0.0300  &1.45 &1.20 &2.70 &0.36 &Pseudo &Y\\
  98 &$J143411.25+003656.6\_1850$ &218.5458 & -0.6158  &0.0299   & 1.70 & 0.55 & 1.80 & 0.22  &Pseudo &Y\\
 99 &$J143519.72+002832.6\_1854$ &218.8292 & -0.4769  &0.0295  & 1.00 & 0.35 & 1.32 & 0.00 &Pseudo &Y\\
 100 &$J143540.07+001217.7\_1857$ &218.8662 &0.1043 &0.0293 &1.80 &1.20 &3.70 &0.50 &Pseudo &Y\\
 101 &$J144258.24+001608.3\_1895$ &220.7721 &0.3184 &0.0291 &1.30 &0.65 &1.78 &0.43 &Pseudo &Y\\
 105 &$J144503.29+003137.1\_1907$ &221.2625 & -0.5269  &0.0289 & 1.70 & 0.60 & 5.30 & 0.01 &Pseudo & \\
 106 &$J144613.34+005157.7\_1912$ &221.5542 & -0.8661  &0.0289  & -- & -- & 1.76 & --  &Bulgeless & \\
 107 &$J145052.33+010956.4\_1934$ &222.7167 &  1.1656  &0.0272  & 1.50 & 0.67 & 1.30 & 0.20  &Pseudo & \\
 108 &$J150648.62+005124.6\_2020$ &226.7042 & -0.8569  &0.0296  & -- & -- & 0.80 & -- &Bulgeless & \\
 109 &$J151926.88-005526.0\_2114$ &229.8625 & -0.9239  &0.0297  & -- & -- & 0.98 & -- &Bulgeless &\\
 110 &$J152024.55-001330.8\_2118$ &230.1000 & -0.2253  &0.0287 & -- & -- & 1.10 & -- &Bulgeless & \\
 112 &$J152254.81-005343.2\_2128$ &230.7292 & -0.8953  &0.0280  & 1.28 & 0.41 & 2.50 & 0.04 &Pseudo &Y\\
 113 &$J152327.26-010956.0\_2132$ &230.8905 &-1.2153 &0.0277 &2.00 &2.17 &5.20 &0.33 &Classical &\\
 116 &$J153117.25+000735.4\_2183$ &232.8208 & -0.1264  &0.0293 & 1.55 & 0.75 & 3.60 & 0.08 &Pseudo &Y\\
\hline
\hline
\label{table_delgado2}
\end{tabular}
\caption{Results of 1D fitting (2/2) for the 66 spiral and S0 galaxies of the \citet{Delgado2010} sample.}
\end{table}
\end{landscape}

\section[]{Impact of the softening value and of the resolution on results.}
\label{appen:soft_and_res}

In this section, we describe the impact of the value of the softening radius and of the resolution on results for three different orbits.

\begin{figure}
\centering
\includegraphics[width=0.35\textwidth]{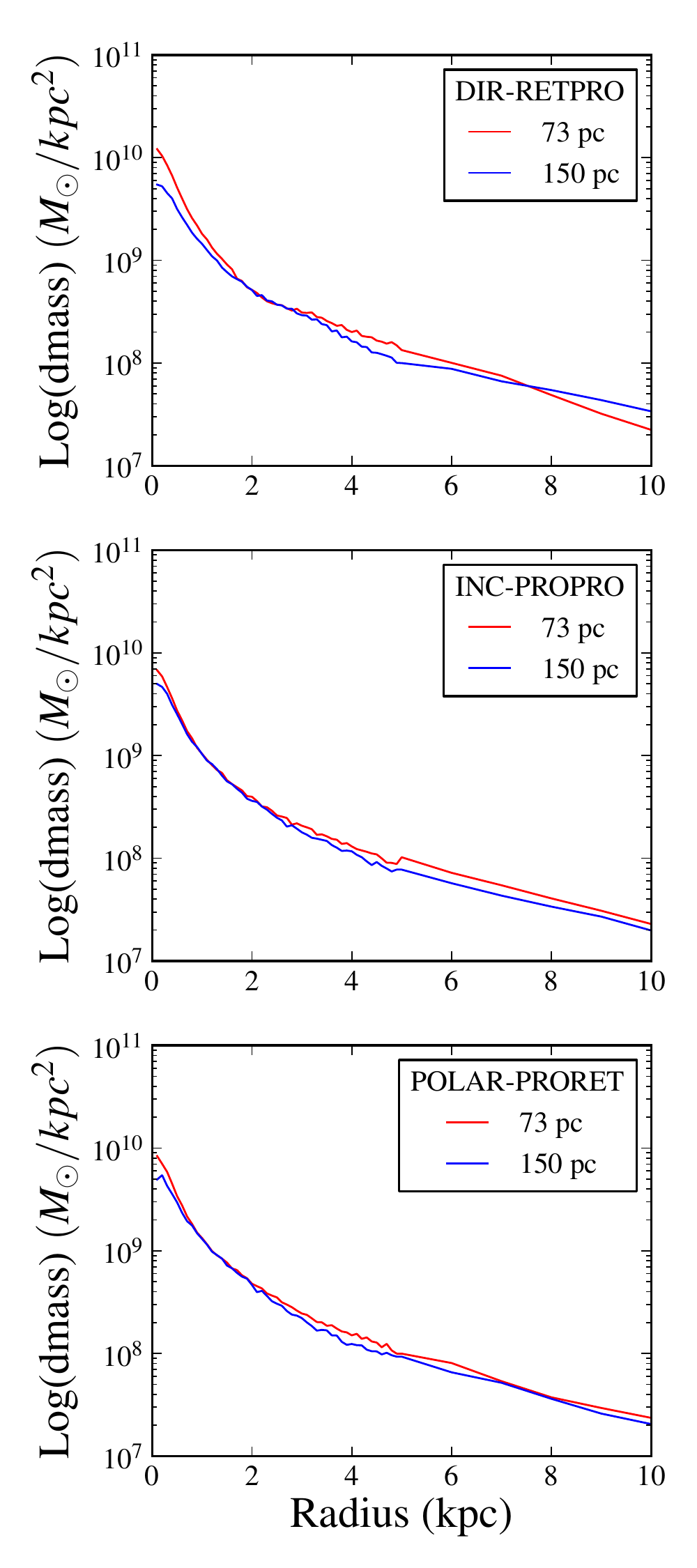} \\  \bigskip \includegraphics[width=0.3\textwidth]{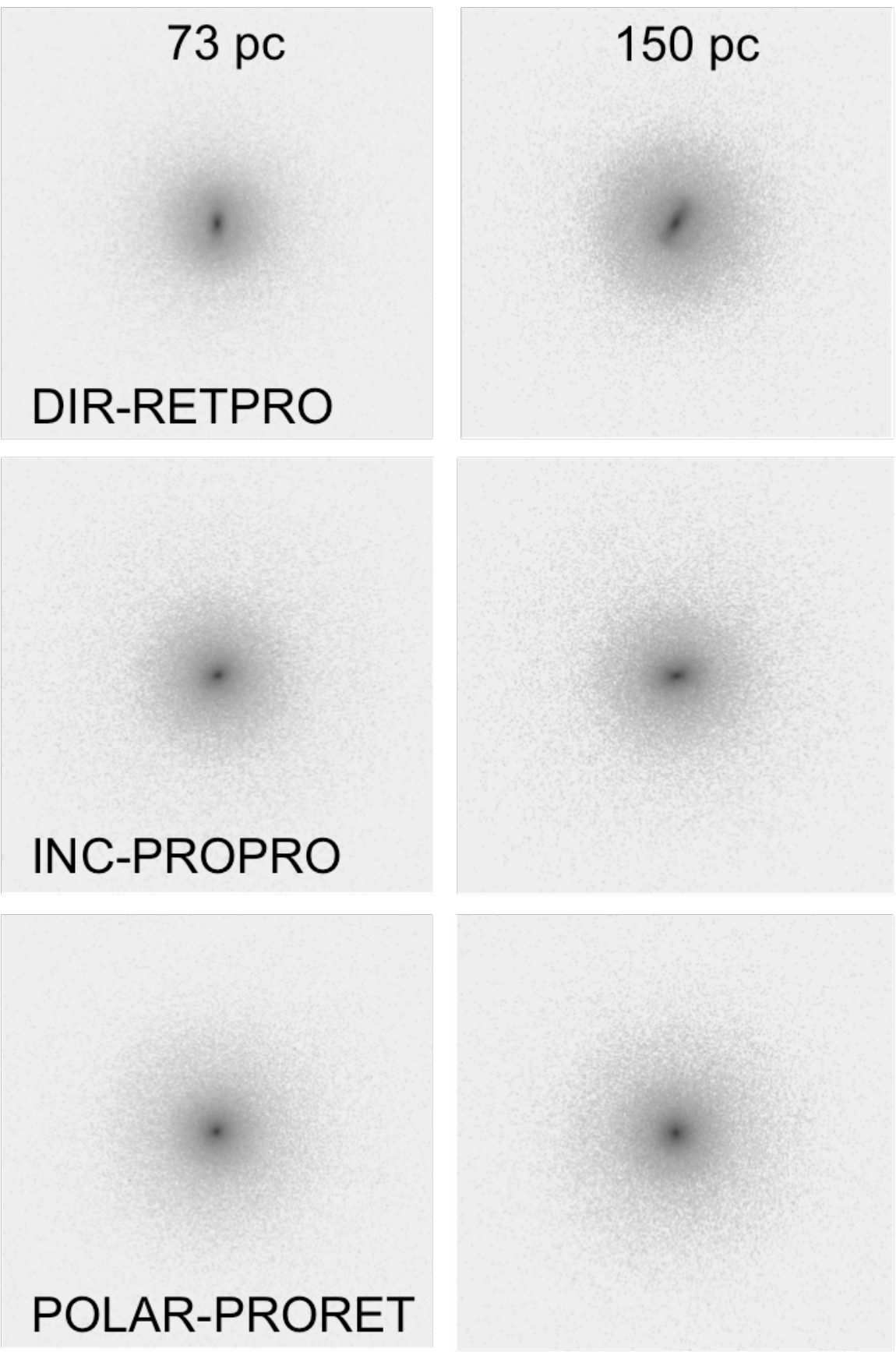} 
  \caption{{\it Top three panels }: Superimposition of mass profile for three simulations with a softening radius of 73 pc (in red in the online version) and 150 pc (in blue in the online version). {\it Bottom panels }: corresponding 2D images within boxes of 60 kpc.}
\label{Fig_softening}
	\end{figure}

The softening radius flattened the structures below 2.8 times this radius in the center  \citep{Springel2005} and impacts the bar formation. However, the value of 73 pc is consistent with the values taken by \citet{Peschken2017}, \citet{Athanassoula2016} or by \citet{Cox2006}, i.e., between 25 and 100 pc. We have tested another value of softening : 150 pc. The effect of increasing the softening radius on the profiles of three different orbits can be seen in Figure \ref{Fig_softening}. The first orbit (DIR-RETPRO) is an orbit favorable to resonances, and the change of softening may affect the profile center. For the other two cases without bars we can see the flattening at the center althought the mass profile is generally unaffected by the softening choice. 

\begin{figure}
\centering
\includegraphics[width=0.55\textwidth]{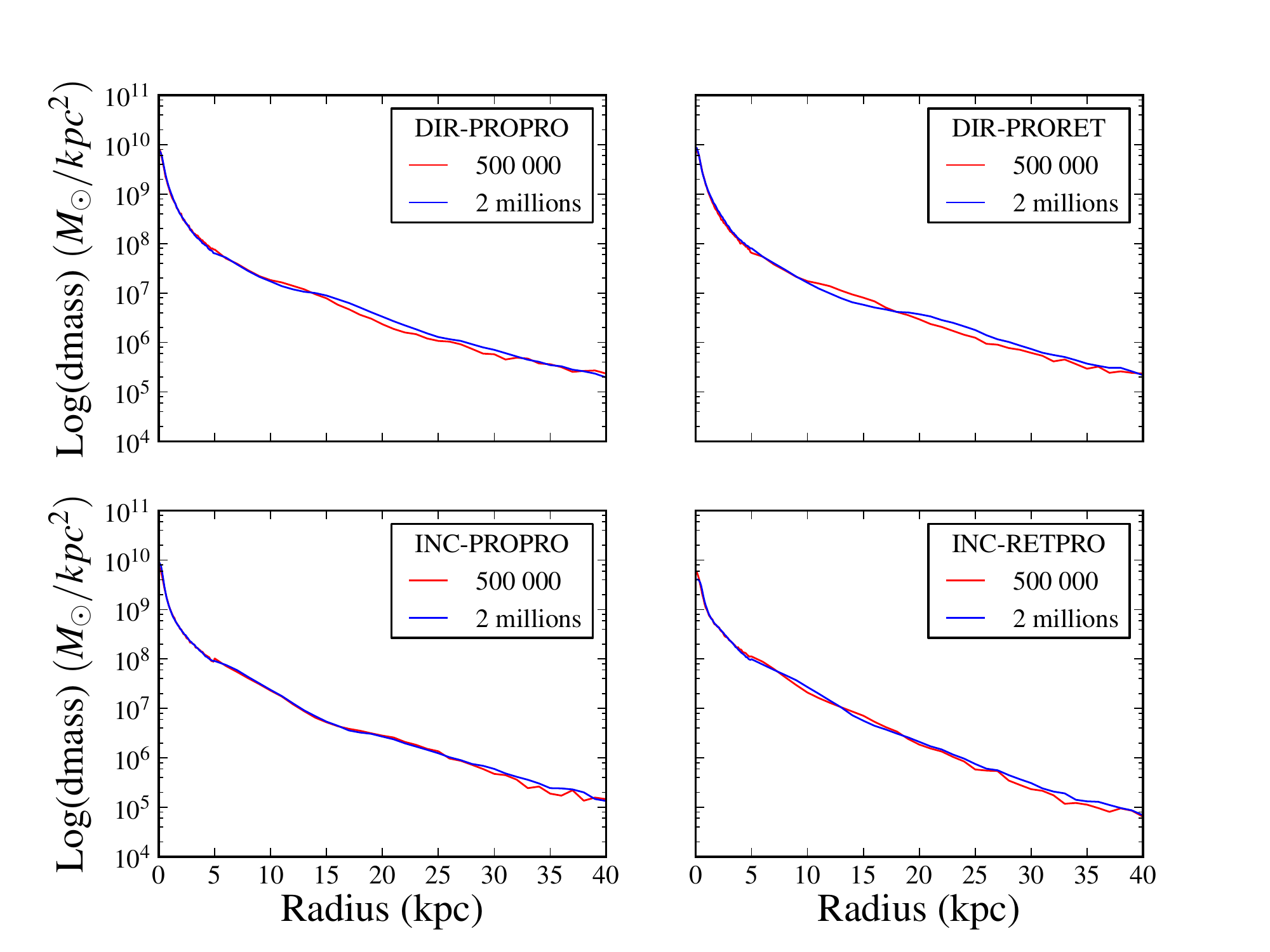}
  \caption{Superimposition of mass profiles for the 4 simulations with 500 000 particles (in red in the online version) and 2 millions of particules (in blue in the online version).}
\label{Fig_resolution}
	\end{figure}

\begin{figure}
\centering
\includegraphics[width=0.5\textwidth]{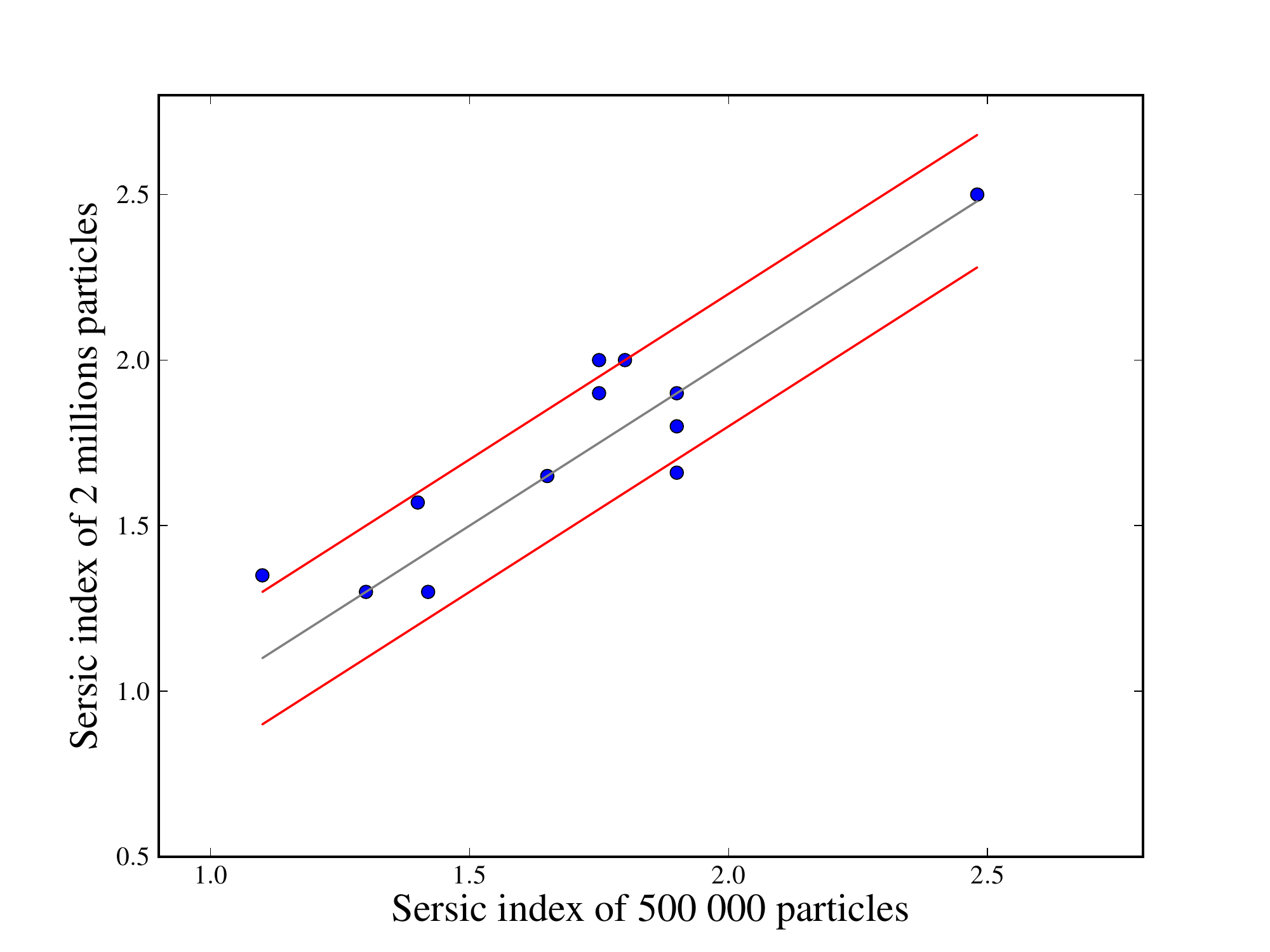}
  \caption{Comparison between the Sersic index of simulations for 500 000 particles and for 2 millions of particles. The grey line is the values if Sersic index was that of 500 000 for both sets of simulations. The red lines (in the online version) are the grey line $\pm 0.2$, which corresponds to the uncertainty estimated on the Sersic index.}
\label{Fig_resolution2}
	\end{figure}

In Figure \ref{Fig_resolution}, we can see the superposition of mass profile for 4 simulations with 500 000 particles and 2 millions of particules. We conclude that the number of particles does not change much the mass profile and the analysis. We can see the comparison of Sersic index between 500 000 particles and 2 millions of particles on Figure \ref{Fig_resolution2}. The only difference is that 2 millions of particules models help to generate resonance effects and we find 3 more bars than for 500 000 particules models (see Table \ref{Table_fit}).

\section[]{Comparison between photometric B/T ratio and B/T ratio from angular momentum decomposition.}
\label{appen:bt}

\begin{figure}
\centering
\includegraphics[width=0.5\textwidth]{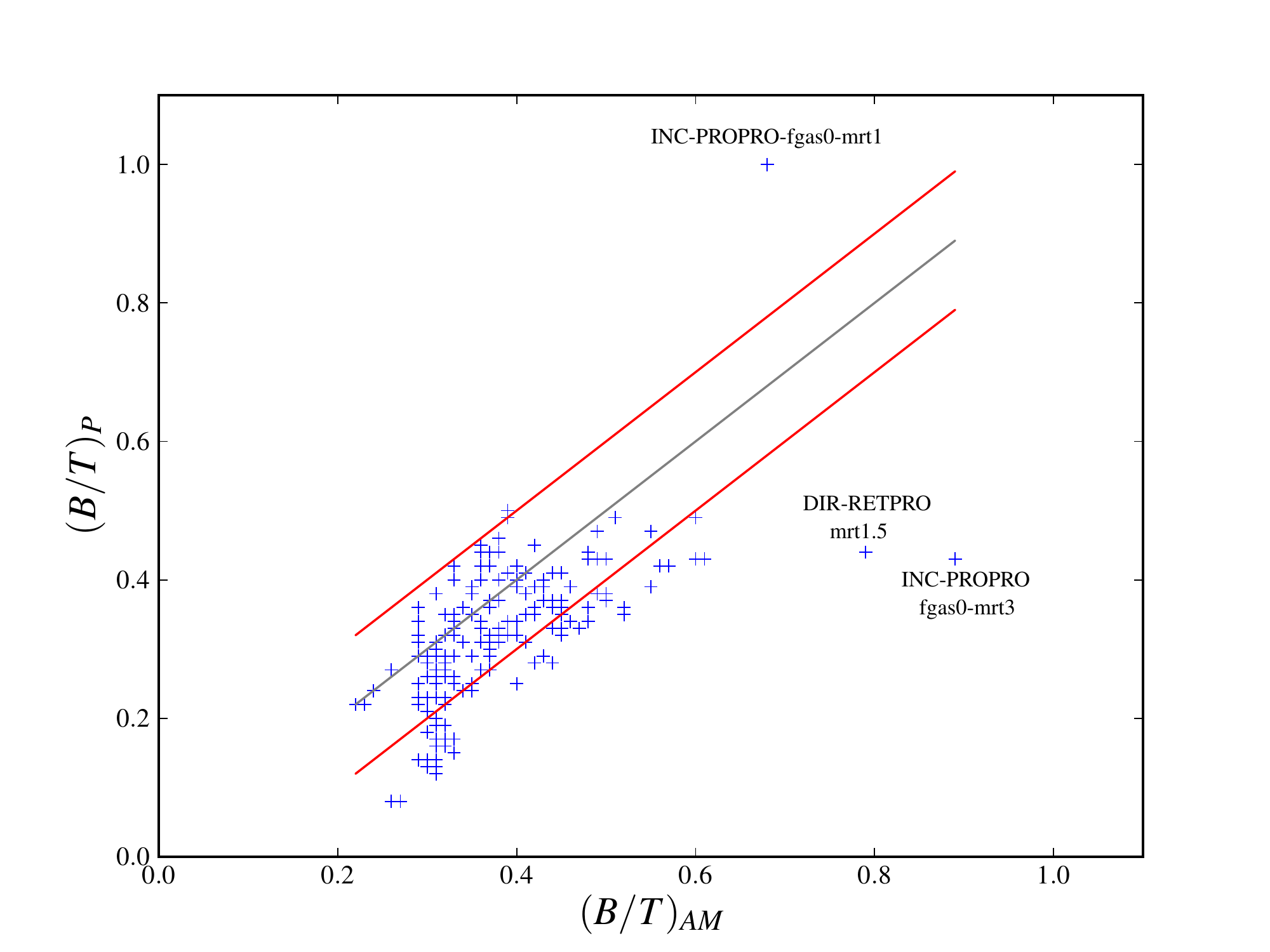}
  \caption{Comparison between the photometric B/T ratio and the B/T ratio from angular momentum decomposition for all the 182 simulations analyzed. The grey line is the values if B/T ratio was equal for both methods. The red lines (in the online version) are the grey line $\pm 0.1$, which corresponds to the uncertainty estimated on the measured photometric B/T ratios.}
\label{Fig_bt}
	\end{figure}

Figure \ref{Fig_bt} shows the distribution of the photometric B/T ratio values in function to the B/T ratio from angular momentum decomposition. The error estimated for the photometric B/T ratio is $\pm 0.1$. The only parameter involved in the measure of B/T ratio from angular momentum is the radius chosen to have velocity of stars inside this radius, this could also give a small error that we could add to the previous one. Consider this, only three points are really out of uncertainty, which are almost all elliptical or S0 galaxies : 

\begin{itemize}
\item INC-PROPRO-fgas0-mrt1 : the simulation INC-PROPRO without gas and with a mass ratio 1:1, which is an elliptical galaxy. \\
\item INC-PROPRO-fgas0-mrt3 :  the simulation INC-PROPRO without gas and with a mass ratio 3:1, which is an object with a small bar and a large thick disk.\\
\item DIR-RETPRO-mrt1.5 : the simulation DIR-RETPRO with a mass ratio of 1.5:1, which has a large thick disk.\\
\end{itemize}	
 
\section[]{Result of simulation fits.}

The following Tables summarize all results of the disk+bulge decomposition from the simulations. The columns are : name of the orbit, bar size in kpc, 5 fitted parameters [$I_e$, $n$, $R_e$, $I_d$, $h$], (B/T) ratios (from photometry and from angular momentum, respectively), gas fraction of the remnant galaxy and mass in $M_{\sun}$. The name of the simulation set is indicated at the top of each panel.  Very compact bulges with $R_{e}$ similar or less than 0.20 kpc (which corresponds to 2.8 times the value of the softening) are unlikely reliable.

\begin{landscape}
\begin{table}
\begin{tabular}{lccccccccccc}
\hline
\hline
Orbits   &Bar size (kpc) &$I_e$   &n &$R_e$  &$I_d$ &h  &$(B/T)_P$  &$(B/T)_{AM}$  &$f_{gas}$   &$M_b$ ($\times10^{10} M_{\sun}$ )   &Bulge type     \\     
  &$\pm 0.3$ &$\pm 0.20$   &$\pm 0.20$ &$\pm 0.10$  &$\pm 0.20$ &$\pm 0.30$  &$\pm 0.10$  & &  &    &and (comments)   \\ 
\hline
\hline
DIR-PROPRO   &  &10.34 &1.65 &1.00 &8.54 &3.10 &0.36 &0.29 &0.165 &3.99    &Pseudo \\
DIR-PRORET  & &10.56 &1.90 &1.10 &8.48 &3.30 &0.42 &0.33  &0.156 &4.16    &Pseudo         \\
DIR-RETPRO  &4.0  &10.30 &1.10 &0.67 &8.95 &2.70 &0.24 &0.34  &0.048 &5.47    &Pseudo (bar)   \\
DIR-RETRET  &10.0  &10.48 &1.42 &0.91 &8.68 &3.30 &0.39  &0.34 & 0.053  &5.36    &Pseudo (bar, ring)          \\
INC-PROPRO  &0.6  &10.40 &1.80 &1.00 &8.60 &3.40 &0.26 &0.30  &0.177 & 4.30    & Pseudo\\
INC-PRORET   &1.0  &10.40 &1.75 &1.10 &8.58 &3.39 &0.33 &0.36    &0.171  &4.60    &Pseudo      \\
INC-RETPRO  &2.0 &10.81 &2.48 &1.20 &8.70 &3.10 &0.25 &0.29   &0.172  &4.58    &Classical\\
INC-RETRET    &2.8   &10.40 &1.30 &0.43 &9.11 &1.85 &0.14 &0.31    &0.174 &4.76    &Pseudo (bar, ring)    \\
POLAR-PROPRO  &0.4  &10.50 &1.90 &1.08 &8.56 &3.50 &0.31 &0.31  & 0.167 &4.57    &Pseudo    \\
POLAR-PRORET  &0.9  &10.42 &1.75 &1.00 &8.80 &2.70 &0.29 &0.33 &0.157  &4.95    &Pseudo      \\    
POLAR-RETPRO  &1.1  &10.55 &1.40 &0.45 &9.09 &1.90 &0.18 &0.30  &0.143  &4.90    &Pseudo       \\
POLAR-RETRET &   &10.52 &1.90 &1.27 &8.45 &4.00 &0.39 &0.35     &0.174 &5.05    &Pseudo    \\

\hline
2 millions of particles\\
DIR-PROPRO &  &10.34 &1.65 &1.03 &8.49 &3.16 &0.40 &0.33 &0.146 &3.82    &Pseudo\\
DIR-PRORET &2.0  &10.50 &1.80 &1.15 &8.47 &3.40 &0.45 &0.36 &0.126 &4.16    &Pseudo\\
DIR-RETPRO &4.3  &10.40 &1.35 &0.65 &8.91 &2.80 &0.20 &0.31 &0.080 &5.28    &Pseudo (bar)\\
DIR-RETRET &4.0  &10.37 &1.30 &0.60 &8.98 &2.60 &0.17 &0.31 &0.072 &5.15    &Pseudo (bar, ring)\\
INC-PROPRO &1.7   &10.72 &2.00 &0.89 &8.55 &3.60 &0.29 &0.31 &0.156 &4.33    &Classical\\
INC-PRORET &0.6  &10.54 &1.90 &1.05 &8.53 &3.60 &0.32 &0.39 &0.160 &4.48    &Pseudo\\
INC-RETPRO &1.7  &10.90 &2.50 &1.22 &8.60 &3.50 &0.29 &0.29 &0.153 &4.55    &Classical\\
INC-RETRET &2.6  &10.32 &1.30 &0.52 &9.05 &2.00 &0.17 &0.31 &0.151 &4.60    &Pseudo\\
POLAR-PROPRO &0.7 &10.52 &1.90 &1.05 &8.52 &3.52 &0.32 &0.33 &0.154 &4.31    &Pseudo\\
POLAR-PRORET &1.6 &10.56 &2.00 &1.30 &8.49 &3.60 &0.41 &0.39 &0.142 &4.62    &Classical\\
POLAR-RETPRO  &2.5   &10.56 &1.57 &0.55 &8.95 &2.50 &0.16 &0.32 &0.148 &4.95    &Pseudo (bar)\\
POLAR-RETRET &  &10.46 &1.66 &0.87 &8.84 &2.60 &0.28 &0.42 &0.147 &4.90    &Pseudo\\
\hline
fgas : 0.10 - 0.07\\
DIR-PROPRO &1.4  &10.60 &1.55 &0.92 &8.60 &3.50 &0.42 &0.37 &0.023 &5.13    &Pseudo (S0) \\
DIR-PRORET  &1.4  &11.00 &2.00 &1.00 &8.44 &4.10 &0.49 &0.39 &0.026 &5.21   &Classical (large thick disk) \\
DIR-RETPRO  &1.5  &10.84 &1.90 &1.00 &8.46 &4.90 &0.36 &0.52 &0.034 &5.75    &Pseudo (large thick disk) \\
DIR-RETRET  &1.0  &11.05 &2.20 &1.14 &8.30 &5.60 &0.42  &0.57 &0.021 &5.64    &Classical \\
INC-PROPRO   &1.8  &10.61 &1.53 &0.90 &8.53 &4.30 &0.37 &0.43 &0.030 &5.53    &Pseudo (large thick disk)\\
INC-PRORET &1.2   &10.57 &1.46 &0.85 &8.58 &4.10 &0.34 &0.46 &0.033 &5.60    &Pseudo \\
INC-RETPRO &1.2  &10.85 &1.90 &0.96 &8.53 &4.50 &0.34 &0.48 &0.026 &5.63    &Pseudo (large thick disk) \\
INC-RETRET  &0.3 &10.97 &2.06 &1.02 &8.47 &4.70 &0.38  &0.49 &0.026 &5.63    &Classical (large thick disk) \\
POLAR-PROPRO &1.5 &10.70 &1.65 &0.90 &8.49 &4.60 &0.35 &0.45 &0.025 &5.49    &Pseudo (large thick disk) \\
POLAR-PRORET  &1.0  &11.47 &2.80 &1.35 &8.23 &5.90 &0.47 &0.49 &0.027 &5.53    & Classical (large thick disk)\\
POLAR-RETPRO   &1.8   &11.63 &2.90 &1.19 &8.35 &5.30 &0.43 &0.49 &0.034 &5.66   &Classical (large thick disk) \\
POLAR-RETRET  &1.2  &11.92 &3.25 &1.30 &8.20 &6.00 &0.49 &0.51 &0.018 &5.54    &Classical (large thick disk)  \\
\hline
\hline
\end{tabular}
\label{Table_fit}
\caption{Results of 1D fitting (1/7) for the simulated galaxies.}
\end{table}
\end{landscape}

\newpage
\begin{landscape}
\begin{table}
\begin{tabular}{lccccccccccc}
\hline
\hline
Orbits   &Bar size (kpc) &$I_e$   &n &$R_e$  &$I_d$ &h  &$(B/T)_P$  &$(B/T)_{AM}$  &$f_{gas}$   &$M_b$ ($\times10^{10} M_{\sun}$ )   &Bulge type     \\     
  &$\pm 0.3$ &$\pm 0.20$   &$\pm 0.20$ &$\pm 0.10$  &$\pm 0.20$ &$\pm 0.30$  &$\pm 0.10$  & &  &    &and (comments)   \\ 
\hline
\hline
fgas : 0.36 - 0.26\\
DIR-PROPRO &1.7  &10.59 &1.57 &0.80 &8.69 &3.00 &0.36 &0.37 &0.101 &4.79    &Pseudo (S0) \\
DIR-PRORET &  &10.55 &1.60 &0.94 &8.64 &3.10 &0.42 &0.40 &0.088 &4.90    &Pseudo \\
DIR-RETPRO &1.3  &10.50 &1.40 &0.77 &8.65 &3.85 &0.28 &0.44 &0.085 &5.57    &Pseudo \\
DIR-RETRET  &0.7  &10.88 &1.90 &0.87 &8.60 &4.00 &0.33  &0.45 &0.071 &5.62    &Pseudo \\
INC-PROPRO &1.7  &10.92 &1.95 &0.85 &8.54 &3.80 &0.37  &0.38 &0.108 &5.14    &Pseudo\\
INC-PRORET    &1.7  &10.43 &1.40 &0.87 &8.65 &3.50 &0.34 &0.36    &0.171  &4.60    &Pseudo      \\
INC-RETPRO &1.1  &10.84 &2.00 &1.00 &8.55 &3.90 &0.36 &0.44 &0.091 &5.18    &Classical \\
INC-RETRET  &  &10.80 &1.80 &0.87 &8.55 &3.90 &0.37 &0.45 &0.099 &5.28    &Pseudo (thick disk) \\
POLAR-PROPRO  &1.3  &10.70 &1.75 &0.93 &8.50 &4.00 &0.38 &0.41 &0.099 &5.14    &Pseudo (S0) \\
POLAR-PRORET  &2.5  &10.50 &1.50 &0.94 &8.61 &3.50 &0.39 &0.46 &0.091 &5.41    &Pseudo (bar)\\
POLAR-RETPRO  &1.1 &10.61 &1.55 &0.81 &8.67 &3.50 &0.33 &0.47 &0.104 &5.45    &Pseudo \\
POLAR-RETRET  &0.9  &10.47 &1.39 &0.84 &8.70 &3.10 &0.38 &0.50 &0.091 &5.36    &Pseudo (thick disk) \\
\hline
fgas : 0.92 - 0.72\\
DIR-PROPRO  &2.3 & 9.90 &1.32 &1.13 &8.34 &3.50 &0.38 &0.31 &0.197 &3.35   &Pseudo (bar) \\
DIR-PRORET &5.0  &10.10 &1.40 &1.10 &8.40 &3.20 &0.46 &0.38 &0.161 &3.54   &Pseudo (bar) \\
DIR-RETPRO  &8.2  &10.15 &1.17 &0.85 &8.80 &3.50 &0.22 &0.29 &0.047 &5.19   &Pseudo (bar, ring) \\
DIR-RETRET  &4.3  &10.10 &1.10 &0.80 &9.00 &2.50 &0.23  &0.32 &0.068 &5.35   &Pseudo (bar) \\
INC-PROPRO &3.5 & 9.47 &1.00 &1.08 &8.38 &3.80 &0.22 &0.22 &0.232 &3.69   &Pseudo (bar)\\
INC-PRORET &1.3 & 9.45 &1.00 &1.29 &8.55 &3.30 &0.26  &0.33 &0.218 &3.99   &Pseudo \\
INC-RETPRO &4.0  & 9.45 &1.05 &1.36 &8.50 &3.65 &0.24 &0.24 &0.225 &4.04   &Pseudo (bar, ring) \\
INC-RETRET  &3.1  & 9.50 &1.20 &1.45 &8.44 &4.10 &0.22 &0.23 &0.217 &4.19   &Pseudo (bar, ring) \\
POLAR-PROPRO &3.9  & 9.64 &1.20 &1.42 &8.44 &3.50 &0.34  &0.29 &0.201 &3.90   &Pseudo (bar, ring) \\
POLAR-PRORET &1.9  & 9.68 &1.10 &1.23 &8.52 &3.90 &0.25  &0.33 &0.199 &4.36   &Pseudo\\
POLAR-RETPRO  &4.0  & 9.74 &1.30 &1.35 &8.50 &3.80 &0.27 &0.26 &0.199 &4.67   &Pseudo (bar, ring) \\
POLAR-RETRET &3.2  & 9.62 &1.00 &1.45 &8.42 &3.80 &0.40 &0.38 &0.194 &4.58   &Pseudo (S0, bar) \\
\hline
Mass progenitors : 0.5M\\
DIR-RETPRO &7.0  &10.31 &1.40 &0.86 &8.38 &3.00 &0.49 &0.60  &0.016 &2.60   &Pseudo (bar, ring) \\
INC-PROPRO &  & 9.70 &1.40 &1.15 &8.25 &3.30 &0.32 &0.37 &0.241 &2.22   &Pseudo \\
POLAR-PRORET &0.4  & 9.62 &1.30 &1.27 &8.33 &3.10 &0.35 &0.42 &0.202 &2.38   &Pseudo\\
\hline
Mass progenitors : 0.5M-2millions\\
DIR-RETPRO   &4.3  & 9.67 &1.50 &1.28 &8.37 &3.20 &0.27  &0.31 &0.061 &2.36   &Pseudo (bar, ring)\\
INC-PROPRO &4.3  & 9.65 &1.40 &1.30 &8.15 &3.60 &0.37  &0.37  &0.224 &2.21   &Pseudo (bar)\\
POLAR-PRORET   &3.2  & 9.57 &1.20 &1.20 &8.32 &3.15 &0.34 &0.39 &0.216 &2.38   &Pseudo (bar)\\
\hline
Mass progenitors : 2M\\
DIR-RETPRO &0.3  &10.60 &1.55 &1.05 &8.90 &3.80 &0.29  &0.35 &0.091 &10.3 &Pseudo\\
INC-PROPRO   &1.9 &10.40 &1.30 &1.00 &8.77 &4.00 &0.31 &0.36 &0.106 &8.63   &Pseudo\\
POLAR-PRORET  &0.4  &10.85 &2.10 &1.60 &8.56 &4.80 &0.45 &0.42 &0.099 &  9.42   & Classical (disk ring)\\
\hline
\hline
\end{tabular}
\caption{Results of 1D fitting (2/7) for the simulated galaxies.}
\end{table}
\end{landscape}

\newpage
\begin{landscape}
\begin{table}
\begin{tabular}{lccccccccccc}
\hline
\hline
Orbits   &Bar size (kpc) &$I_e$   &n &$R_e$  &$I_d$ &h  &$(B/T)_P$  &$(B/T)_{AM}$  &$f_{gas}$   &$M_b$ ($\times10^{10} M_{\sun}$ )   &Bulge type     \\     
  &$\pm 0.3$ &$\pm 0.20$   &$\pm 0.20$ &$\pm 0.10$  &$\pm 0.20$ &$\pm 0.30$  &$\pm 0.10$  & &  &    &and (comments)   \\ 
\hline
\hline
Mass ratio : 1.5:1\\
DIR-RETPRO  &0.3  &10.25 &1.15 &1.00 &8.98 &2.25 &0.44  &0.79 &0.034 &5.87   &Pseudo\\
INC-PROPRO  &  &10.73 &2.00 &1.05 &8.77 &2.80 &0.36 &0.42 &0.137 &5.19   &Classical\\
POLAR-PRORET &  &10.30 &1.40 &1.12 &8.73 &3.00 &0.42 &0.56  &0.136 &5.85   &Pseudo (S0)\\
\hline
Mass ratio : 4.5:1\\
DIR-RETPRO &8.0   &10.45 &1.50 &0.85 &8.44 &4.41 &0.31 &0.38 &0.164 &5.08   &Pseudo (bar, ring)\\
INC-PROPRO  &1.8  &10.62 &2.03 &1.00 &8.40 &4.10 &0.29 &0.30 &0.192 &4.07   &Classical (bar)\\
POLAR-PRORET &  &10.70 &2.20 &1.22 &8.53 &3.55 &0.35 &0.33 &0.170 &4.48   &Classical\\
\hline
Gas extension : 9 kpc\\
DIR-RETPRO  &3.9   &10.46 &1.50 &0.89 &8.61 &4.50 &0.24 &0.34 &0.086 &5.77   &Pseudo (bar, ring)\\
INC-PROPRO &1.8   &10.52 &1.80 &1.00 &8.59 &3.60 &0.30 &0.31 &0.150 &4.83   &Pseudo\\
POLAR-PRORET &1.7  &11.14 &2.75 &1.66 &8.26 &4.70 &0.50  &0.39 &0.152 &5.23   &Classical\\
\hline
Gas extension : 21 kpc\\
DIR-RETPRO  &10.7  &10.57 &1.55 &1.00 &8.49 &3.80 &0.47 &0.55 &0.038 &5.15   &Pseudo (bar)\\
INC-PROPRO &2.4  &10.35 &1.70 &0.95 &8.56 &3.35 &0.28  &0.32 &0.177 &4.07   &Pseudo\\
POLAR-PRORET &3.0  &10.38 &1.90 &1.55 &8.30 &4.50 &0.44  &0.36 &0.161 &4.82   &Pseudo\\
\hline
Pericenter : 8 kpc\\
DIR-RETPRO &10.5 &10.57 &1.60 &1.02 &8.51 &3.90 &0.43 &0.50 &0.098 &5.61   &Pseudo (bar, ring)\\
INC-PROPRO &1.7 &10.18 &1.70 &1.32 &8.45 &4.00 &0.32 &0.29 &0.171 &4.51   &Pseudo\\
POLAR-PRORET &3.4  &11.20 &2.80 &1.30 &8.65 &3.50 &0.31 &0.31 &0.152 &5.16   &Classical\\
\hline
Pericenter : 24 kpc\\
DIR-RETPRO &5.0 &10.40 &1.35 &0.90 &8.62 &4.00 &0.32 &0.38 &0.066 &5.44   &Pseudo (bar, disk ring)\\
INC-PROPRO &2.2 &10.11 &1.48 &1.20 &8.45 &4.00 &0.33 &0.33 &0.170 &4.43   &Pseudo\\
POLAR-PRORET &1.7 &10.55 &2.00 &1.35 &8.36 &4.50 &0.39 &0.40 &0.164 &4.79   &Classical \\
\hline
Softening : 150 pc\\
DIR-RETPRO  &4.1  &10.20 &1.60 &1.30 &8.46 &4.50 &0.31 &0.34 &0.165 &5.54   &Pseudo (bar, ring)\\
INC-PROPRO &2.4  &10.20 &1.69 &1.21 &8.41 &3.70 &0.35 &0.35 &0.209 &3.96   &Pseudo (bar)\\
POLAR-PRORET  &   &10.00 &1.40 &1.20 &8.58 &3.23 &0.34 &0.40 &0.191 &4.63   &Pseudo\\
\hline
Feedback : median\\
DIR-RETPRO &3.2  &10.68 &1.70 &0.98 &8.38 &4.50 &0.43 &0.48 &0.106 &5.15   &Pseudo (bar, ring)\\
INC-PROPRO  &1.2  &10.80 &2.20 &1.00 &8.60 &3.70 &0.26 &0.31 &0.126 &4.50   &Classical\\
POLAR-PRORET &  &11.15 &2.15 &0.90 &8.55 &3.50 &0.47 &0.49 &0.096 &5.15   &Classical (S0)\\
\hline
Feedback : 2.5*median\\
DIR-RETPRO  &3.2  &10.48 &1.54 &0.85 &8.60 &3.50 &0.32 &0.35 &0.114 &5.08   &Pseudo (bar, ring)\\
INC-PROPRO &  &10.70 &2.15 &1.05 &8.51 &4.10 &0.25 &0.31 &0.154 &4.46   &Classical\\
POLAR-PRORET &  &10.60 &1.90 &1.20 &8.47 &4.10 &0.39 &0.42 &0.133 &5.18   &Pseudo \\
\hline
\hline
\end{tabular}
\caption{Results of 1D fitting (3/7) for the simulated galaxies.}
\end{table}
\end{landscape}

\newpage
\begin{landscape}
\begin{table}
\begin{tabular}{lccccccccccc}
\hline
\hline
Orbits   &Bar size (kpc) &$I_e$   &n &$R_e$  &$I_d$ &h  &$(B/T)_P$  &$(B/T)_{AM}$  &$f_{gas}$   &$M_b$ ($\times10^{10} M_{\sun}$ )   &Bulge type     \\     
  &$\pm 0.3$ &$\pm 0.20$   &$\pm 0.20$ &$\pm 0.10$  &$\pm 0.20$ &$\pm 0.30$  &$\pm 0.10$  & &  &    &and (comments)   \\  
\hline
\hline
Time evol : DIR-PROPRO\\
0.8 Gyr  & &10.93 &2.50 &1.04 &8.58 &2.90 &0.32 &0.32 &0.251 &4.02 &Classical \\
2.2 Gyr  & &10.78 &2.10 &0.91 &8.55 &3.1 &0.35 &0.32   &0.211  &4.00 &Classical\\
3.5 Gyr  &  &10.60 &2.00 &0.98 &8.55 &3.10 &0.33 &0.33   &0.186  &3.99 &Classical\\
4.9 Gyr & &10.34 &1.70 &1.00 &8.54 &3.10 &0.34 &0.33   &0.173  &3.99 &Pseudo \\
6.3 Gyr  &  &10.34 &1.65 &1.00 &8.54 &3.10 &0.36 &0.29 &0.165 &3.99 &Pseudo\\
7.7 Gyr  & &10.34 &1.65 &1.00 &8.55 &3.10 &0.36 &0.34   &0.159  &3.98 &Pseudo\\
9.0 Gyr &  &10.34 &1.65 &1.00 &8.54 &3.10 &0.36 &0.34   &0.156  &3.98 &Pseudo\\
\hline
Time evol : DIR-PRORET\\
0.8 Gyr &1.7   &10.80 &2.30 &1.12 &8.47 &3.30 &0.38 &0.35   &0.257   &4.23 &Classical\\
2.2 Gyr &1.7  &10.80 &2.10 &1.00 &8.47 &3.30 &0.42 &0.36   &0.209  &4.21&Classical\\
3.5 Gyr &  &10.72 &2.00 &1.00 &8.47 &3.30 &0.42 &0.36   &0.181  &4.17 &Classical\\
4.9 Gyr &  &10.63 &1.90 &1.00 &8.47 &3.30 &0.42 &0.36   &0.165  &4.17 &Pseudo\\
6.3 Gyr & &10.56 &1.90 &1.10 &8.48 &3.30 &0.42 &0.33  &0.156 &4.16   &Pseudo\\
7.7 Gyr &  &10.62 &1.90 &1.05 &8.47 &3.30 &0.44 &0.37   &0.149  &4.15 &Pseudo\\
9.0 Gyr &  &10.62 &1.90 &1.05 &8.47 &3.30 &0.44 &0.38   &0.146  &4.15 &Pseudo\\
\hline
Time evol : DIR-RETPRO\\
0.8 Gyr &2.8  &10.20 &1.00 &0.47 &9.04 &2.40 &0.08 &0.26   &0.205  &5.44 &Pseudo (bar)\\
2.2 Gyr &4.7  &10.24 &1.10 &0.60 &8.96 &2.60 &0.19 &0.32   &0.125  &5.46 &Pseudo (bar)\\
3.5 Gyr &4.5  &10.24 &1.10 &0.65 &8.96 &2.60 &0.22 &0.32   &0.087  &5.47 &Pseudo (bar)\\
4.9 Gyr &4.1  &10.25 &1.05 &0.66 &8.96 &2.60 &0.24 &0.34   &0.062  &5.47 &Pseudo (bar)\\
6.3 Gyr &4.0  &10.30 &1.10 &0.67 &8.95 &2.70 &0.24 &0.34  &0.048 &5.47   &Pseudo (bar)\\
7.7 Gyr &4.0  &10.26 &1.10 &0.70 &8.95 &2.70 &0.24 &0.35   &0.040  &5.47 &Pseudo (bar)\\
9.0 Gyr &3.7  &10.24 &1.05 &0.70 &8.96 &2.70 &0.25 &0.35   &0.034  &5.46 &Pseudo (bar)\\
\hline
Time evol : DIR-RETRET\\
0.8 Gyr &1.4  &10.05 &1.05 &0.42 &9.15 &2.00 &0.08 &0.27   &0.208  &5.49 &Pseudo (bar)\\
2.2 Gyr &3.7  &10.15 &1.05 &0.49 &9.15 &2.10 &0.12 &0.31   &0.119  &5.47 &Pseudo (bar)\\
3.5 Gyr &5.5  &10.38 &1.35 &0.74 &8.94 &2.60 &0.25 &0.40   &0.086  &5.45 &Pseudo (bar)\\
4.9 Gyr &8.3  &10.45 &1.45 &0.97 &8.64 &3.60 &0.37 &0.50   &0.065  &5.40 &Pseudo (bar)\\
6.3 Gyr  &10.0 &10.48 &1.42 &0.91 &8.68 &3.30 &0.39  &0.55 & 0.053  &5.36   &Pseudo (bar)          \\
7.7 Gyr  &10.0  &10.48 &1.48 &1.02 &8.59 &3.60 &0.43 &0.60   &0.045  &5.35 &Pseudo (bar)\\
9.0 Gyr &9.5  &10.49 &1.45 &0.99 &8.60 &3.60 &0.43 &0.61   &0.039  &5.35 &Pseudo (bar)\\
\hline
\hline
\end{tabular}
\caption{Results of 1D fitting (4/7) for the simulated galaxies.}
\end{table}
\end{landscape}

\newpage
\begin{landscape}
\begin{table}
\begin{tabular}{lccccccccccc}
\hline
\hline
Orbits   &Bar size (kpc) &$I_e$   &n &$R_e$  &$I_d$ &h  &$(B/T)_P$  &$(B/T)_{AM}$  &$f_{gas}$   &$M_b$ ($\times10^{10} M_{\sun}$ )   &Bulge type     \\     
  &$\pm 0.3$ &$\pm 0.20$   &$\pm 0.20$ &$\pm 0.10$  &$\pm 0.20$ &$\pm 0.30$  &$\pm 0.10$  & &  &    &and (comments)   \\ 
\hline
\hline
Time evol : INC-PROPRO\\
0.8 Gyr  &1.7  &11.82 &3.40 &1.00 &8.54 &3.40 &0.31 &0.29   & 0.275 &4.27   &Classical\\
2.2 Gyr &0.9  &10.87 &2.52 &1.18 &8.55 &3.40 &0.29 &0.30 &0.235 &4.32   &Classical\\
3.5 Gyr &0.8  &10.50 &2.00 &1.10 &8.55 &3.40 &0.29 &0.30 &0.205 &4.30   &Classical\\
4.9 Gyr &0.8  &10.70 &2.20 &1.05 &8.57 &3.40 &0.28 &0.30 &0.189 &4.30   &Classical\\
6.3 Gyr &0.6  &10.40 &1.80 &1.00 &8.60 &3.40 &0.26  &0.30 &0.177 & 4.30   &Pseudo\\
7.7 Gyr &0.6  &10.28 &1.65 &1.00 &8.58 &3.40 &0.27 &0.31 &0.168 & 4.29   &Pseudo\\
9.0 Gyr &0.6  &10.27 &1.65 &1.00 &8.59 &3.40 &0.26 &0.31 &0.161 &4.29   &Pseudo\\
\hline
Time evol : INC-PRORET\\
0.8 Gyr   &2.7  &10.70 &2.33 &1.23 &8.54 &3.35 &0.31 &0.41   &0.302  &4.51 &Classical\\
2.2 Gyr   &1.7  &10.50 &2.00 &1.20 &8.52 &3.40 &0.35 &0.41   &0.241  &4.58 &Pseudo\\
3.5 Gyr   &1.8  &10.35 &1.80 &1.22 &8.54 &3.40 &0.35 &0.41   &0.205  &4.58 &Pseudo\\
4.9 Gyr  &1.5  &10.42 &1.80 &1.10 &8.59 &3.34 &0.32 &0.40   &0.189  &4.62 &Pseudo\\
6.3 Gyr    &1.0  &10.40 &1.75 &1.10 &8.58 &3.39 &0.33 &0.36    &0.171  &4.60   &Pseudo      \\
7.7 Gyr &1.3  &10.45 &1.80 &1.10 &8.55 &3.50 &0.34 &0.40   &0.158  &4.59 &Pseudo\\
9.0 Gyr  &1.5  &10.45 &1.80 &1.10 &8.55 &3.50 &0.34 &0.40   &0.149  &4.58 &Pseudo\\ 
\hline
Time evol : INC-RETPRO\\
0.8 Gyr   &1.7  &10.70 &2.35 &1.02 &8.66 &3.10 &0.21 &0.30   &0.278  &4.58 &Classical\\
2.2 Gyr  &1.7  &10.54 &2.05 &0.98 &8.68 &3.10 &0.22  &0.29   &0.239  &4.64 &Classical \\
3.5 Gyr   &1.7 &10.55 &2.10 &1.05 &8.70 &3.10 &0.23 &0.29   &0.208  &4.62 &Classical\\
4.9 Gyr  &2.5  &10.60 &2.10 &1.00 &8.70 &3.10 &0.23 &0.30   &0.188  &4.60 &Classical\\
6.3 Gyr   &2.3 &10.81 &2.48 &1.20 &8.70 &3.10 &0.25 &0.29   &0.172  &4.58   &Classical\\
7.7 Gyr &2.0  &10.69 &2.35 &1.20 &8.70 &3.10 &0.25 &0.29   &0.165  &4.59 &Classical\\
9.0 Gyr  &2.4  &10.55 &2.10 &1.10 &8.70 &3.10 &0.25 &0.29   &0.157  &4.59 &Classical\\
\hline
Time evol : INC-RETRET\\
0.8 Gyr   &1.2  &10.45 &1.17 &0.30 &9.11 &1.65 &0.13 &0.31   &0.285  &4.68 &Pseudo\\
2.2 Gyr   &1.6  &10.40 &1.30 &0.40 &9.11 &1.85 &0.13 &0.30   &0.234  &4.72 &Pseudo\\
3.5 Gyr   &1.7  &10.20 &1.20 &0.45 &9.11 &1.75 &0.14 &0.29   &0.216  &4.79 &Pseudo\\
4.9 Gyr  &2.0  &10.05 &1.00 &0.45 &9.14 &1.70 &0.14 &0.29   &0.191  &4.77 &Pseudo\\
6.3 Gyr  &2.8   &10.40 &1.30 &0.43 &9.11 &1.85 &0.14 &0.31    &0.174 &4.76   &Pseudo     \\
7.7 Gyr &3.0  &10.05 &1.00 &0.47 &9.14 &1.70 &0.14 &0.30   &0.167  &4.77 &Pseudo\\
9.0 Gyr  &3.2  &10.20 &1.20 &0.52 &9.10 &1.85 &0.16 &0.31   &0.156  &4.75 &Pseudo\\
\hline
\hline
\end{tabular}
\caption{Results of 1D fitting (5/7) for the simulated galaxies.}
\end{table}
\end{landscape}

\newpage
\begin{landscape}
\begin{table}
\begin{tabular}{lccccccccccc}
\hline
\hline
Orbits   &Bar size (kpc) &$I_e$   &n &$R_e$  &$I_d$ &h  &$(B/T)_P$  &$(B/T)_{AM}$  &$f_{gas}$   &$M_b$ ($\times10^{10} M_{\sun}$ )   &Bulge type     \\     
  &$\pm 0.3$ &$\pm 0.20$   &$\pm 0.20$ &$\pm 0.10$  &$\pm 0.20$ &$\pm 0.30$  &$\pm 0.10$  & &  &    &and (comments)   \\ 
\hline
\hline
Time evol : POL-PROPRO\\
0.8 Gyr &1.4 &10.46 &1.95 &1.05 &8.56 &3.40 &0.27 &0.32 &0.262 &4.56   &Pseudo\\
2.2 Gyr  &1.1 &10.35 &1.80 &1.07 &8.60 &3.20 &0.29 &0.32 &0.220 &4.58   &Pseudo\\
3.5 Gyr &0.9 &10.28 &1.70 &1.09 &8.55 &3.50 &0.29 &0.32 &0.194 &4.57   &Pseudo\\
4.9 Gyr & &10.33 &1.72 &1.06 &8.60 &3.30 &0.30 &0.31 &0.181 &4.58   &Pseudo\\
6.3 Gyr &  &10.50 &1.90 &1.08 &8.56 &3.50 &0.31 &0.31  & 0.167 &4.57   &Pseudo    \\
7.7  Gyr & &10.14 &1.50 &1.08 &8.58 &3.50 &0.28 &0.32 &0.158 &4.55   &Pseudo\\
9.0 Gyr & &10.00 &1.35 &1.06 &8.58 &3.55 &0.26 &0.32 &0.149 &4.54   &Pseudo\\
\hline
Time evol : POL-PRORET\\
0.8 Gyr  &0.8 &10.44 &1.85 &0.99 &8.77 &2.70 &0.27 &0.37 &0.253 &4.96   &Pseudo\\
2.2 Gyr  &0.8 &10.50 &1.85 &0.99 &8.78 &2.70 &0.29  &0.37 &0.206 &4.95   &Pseudo\\
3.5 Gyr &0.7 &10.45 &1.75 &1.00 &8.75 &2.85 &0.31  &0.37 &0.180 &4.95   &Pseudo\\
4.9 Gyr &0.7 &10.45 &1.75 &0.95 &8.80 &2.75 &0.27 &0.36 &0.168 &4.95   &Pseudo\\
6.3 Gyr &0.9 &10.42 &1.75 &1.00 &8.80 &2.70 &0.29 &0.33 &0.157  &4.95   &Pseudo\\
7.7  Gyr &0.6 &10.40 &1.70 &1.02 &8.69 &3.10 &0.30 &0.37 &0.147 &4.93   &Pseudo\\
9.0 Gyr  &0.0 &10.30 &1.60 &1.00 &8.77 &2.80 &0.29 &0.37 &0.138  &4.91   &Pseudo\\
\hline
Time evol : POL-RETPRO\\
0.8 Gyr   &3.0  &12.22 &2.50 &0.28 &8.93 &2.20 &0.15 &0.33   &0.298  &4.87 &Classical\\
2.2 Gyr   &1.7  &10.85 &1.90 &0.52 &8.99 &2.20 &0.17 &0.33   &0.227  &4.99 &Pseudo\\
3.5 Gyr   &1.2  &10.93 &1.90 &0.50 &9.09 &1.95 &0.19 &0.31   &0.191  &4.99 &Pseudo\\
4.9 Gyr  &0.7  &11.10 &1.95 &0.47 &9.07 &1.95 &0.23 &0.31   &0.160  &4.94 &Pseudo\\
6.3 Gyr &1.1  &10.55 &1.40 &0.45 &9.09 &1.90 &0.18 &0.30  &0.143  &4.90   &Pseudo\\
7.7 Gyr  &2.1  &10.50 &1.50 &0.52 &9.10 &1.95 &0.17 &0.32   &0.139  &4.91 &Pseudo\\
9.0 Gyr  &2.9  &10.55 &1.50 &0.50 &9.09 &1.95 &0.17 &0.33   &0.136  &4.92 &Pseudo\\
\hline
Time evol : POL-RETRET\\
0.8 Gyr   &1.8  &10.50 &1.95 &1.42 &8.24 &4.90 &0.44 &0.48   &0.212  &4.40 &Pseudo\\
2.2 Gyr   &1.1 &10.73 &2.15 &1.22 &8.51 &3.60 &0.39 &0.46   &0.242  &5.05 &Classical\\
3.5 Gyr  &  &10.50 &1.90 &1.30 &8.42 &4.00 &0.41 &0.44   &0.207  &5.04 &Pseudo\\
4.9 Gyr &    &10.40 &1.75 &1.26 &8.43 &4.10 &0.39 &0.43   &0.180  &5.01 &Pseudo\\
6.3 Gyr   & &10.52 &1.90 &1.27 &8.45 &4.00 &0.39  &0.35     &0.174 &5.05   &Pseudo\\
7.7 Gyr &   &10.35 &1.65 &1.25 &8.43 &4.10 &0.40  &0.43 &0.165  &5.05 &Pseudo\\
9.0 Gyr  &   &10.20 &1.52 &1.27 &8.44 &4.20 &0.37 &0.43   &0.153  &5.03 &Pseudo\\
\hline
Mass ratio 1:1 - $F_{gas}$ : 0.2\%-0.2\%  \\
INC-PROPRO &2.0  &13.00 &5.47 &6.50 &0.00 &0.00 &1.00 &0.68 &0.0004 & 7.51   &Classical\\
 & & & & & & & & & & &  (elliptical)      \\
\hline
Mass ratio 3:1 -  $F_{gas}$ : 0.2\%-0.6\%\\
INC-PROPRO  &1.7  &11.60 &2.80 &1.10 &8.55 &4.10 &0.43 &0.89 &0.0006 &5.57   &Classical\\
 & & & & & & & & & & & (large thick disk)\\
\hline
\hline
\end{tabular}
\caption{Results of 1D fitting (6/7) for the simulated galaxies.}
\end{table}
\end{landscape}

\newpage
\begin{landscape}
\begin{table}
\begin{tabular}{lccccccccccc}
\hline
\hline
Orbits   &Bar size (kpc) &$I_e$   &n &$R_e$  &$I_d$ &h  &$(B/T)_P$  &$(B/T)_{AM}$  &$f_{gas}$   &$M_b$ ($\times10^{10} M_{\sun}$ )   &Bulge type     \\     
  &$\pm 0.3$ &$\pm 0.20$   &$\pm 0.20$ &$\pm 0.10$  &$\pm 0.20$ &$\pm 0.30$  &$\pm 0.10$  & &  &    &and (comments)   \\ 
\hline
\hline
fgas : 0.36 - 0.26 - 2.14 Gyr \\
after the first passage\\
DIR-PROPRO  &1.8  &10.60 &1.54 &0.80 &8.66 &3.00 &0.40 &0.36 &0.112 &4.79 &Pseudo \\
DIR-PRORET &1.5  &10.80 &1.82 &0.86 &8.65 &3.10 &0.40 &0.40 &0.107 &4.91 &Pseudo \\
DIR-RETPRO  &1.5  &10.35 &1.30 &0.83 &8.60 &4.00 &0.29 &0.43 &0.130 &5.65 &Pseudo \\
DIR-RETRET    &1.7  &11.30 &2.20 &0.71 &8.60 &4.00 &0.32 &0.45 &0.127 &5.12 &Classical \\
INC-PROPRO  &2.7  &10.75 &1.80 &0.85 &8.54 &3.90 &0.33 &0.38 &0.106 &5.15 &Pseudo\\
INC-PRORET    &2.0  &10.57 &1.50 &0.80 &8.65 &3.50 &0.33 &0.44 &0.099 &5.28 &Pseudo \\
INC-RETPRO  &1.5  &11.05 &2.20 &0.95 &8.55 &3.90 &0.36 &0.45 &0.112 &5.11 &Classical \\
INC-RETRET   &1.7  &11.20 &2.25 &0.85 &8.54 &3.90 &0.37 &0.44 &0.091 &5.41 &Classical \\
POLAR-PROPRO   &2.3  &11.15 &2.20 &0.89 &8.48 &4.00 &0.41 &0.41 &0.132 &5.42 &Classical (S0) \\
POLAR-PRORET   &2.9 &10.95 &1.95 &0.88 &8.60 &3.51 &0.41 &0.45 &0.113 &5.33 &Pseudo\\
POLAR-RETPRO  &1.9  &11.23 &2.10 &0.70 &8.64 &3.50 &0.36 &0.48 &0.123 &5.64 &Classical \\
POLAR-RETRET   &1.0  &10.58 &1.50 &0.80 &8.70 &3.20 &0.35 &0.52 &0.123 &5.25 &Pseudo \\

\hline
\hline
\end{tabular}
\caption{Results of 1D fitting (7/7) for the simulated galaxies.}
\end{table}
\end{landscape}


\bsp	
\label{lastpage}
\end{document}